\documentclass[twocolumn, a4paper, reprint, aps, groupedaddress, superscriptaddress]{revtex4-1}
\pdfoutput=1
\usepackage{graphicx}
\usepackage{dcolumn}
\usepackage{bm}
\usepackage{blindtext}
\usepackage{hyperref}
\usepackage{placeins}
\usepackage{siunitx}
\usepackage{amsmath}
\usepackage{amsfonts}
\usepackage{amssymb}
\usepackage{mathtools}
\usepackage{physics}
\usepackage{tikz}
\usepackage[titletoc]{appendix}
\usepackage{lipsum}
\usepackage{listings}
\usepackage{xcolor}
\usepackage{float}
\usepackage{ulem}
\usepackage[export]{adjustbox}
\usepackage{resizegather}
\usepackage{caption}
\usepackage{color}
\usepackage{subfigure}
\usepackage[bottom]{footmisc}

\DeclareSIUnit \parsec {pc}
\DeclareSIUnit \h {\mbox{$h$}}

\newcommand{\teps}{\tilde \epsilon}
\newcommand{\siteps}{{\rm sign}(\tilde \epsilon)}
\begin{document}

\title{Bayesian model selection on Scalar $\epsilon$-Field Dark Energy}

\author{J. Alberto V\'azquez}
\email{javazquez@icf.unam.mx}
\affiliation{Instituto de Ciencias F\'isicas, Universidad Nacional Aut\'onoma de M\'exico, Cuernavaca, Morelos, 62210, M\'exico}

\author{David Tamayo}
\email{tamayo.ramirez.d.a@gmail.com}
\affiliation{Facultad de Ciencias en F\'isica y Matem\'aticas, Universidad Aut\'onoma de Chiapas, 
Tuxtla Guti\'errez, Chiapas, 29050, M\'exico}
\affiliation{Mesoamerican Centre for Theoretical Physics,
Universidad Aut\'onoma de Chiapas, 
Tuxtla Guti\'errez, Chiapas, 29050, Mexico.}

\author{Anjan A. Sen}
\email{aasen@jmi.ac.in}
\affiliation{Centre for Theoretical Physics, Jamia Millia Islamia, New Delhi-110025, India.}

\author{Israel Quiros}\email{iquiros@fisica.ugto.mx}\affiliation{Departamento Ingenier\'ia Civil, Divisi\'on de Ingenier\'ia, Universidad de Guanajuato, Guanajuato, C.P. 36000, M\'exico.}

\begin{abstract}
The main aim of this paper is to analyse
minimally-coupled scalar-fields  -- quintessence and phantom --  
as the main candidates to explain the accelerated expansion of the universe 
and compare its observables to current cosmological observations; as a byproduct 
we present its python module.
This work includes a parameter $\epsilon$ which allows to incorporate both 
quintessence and phantom fields within the same analysis.
Examples of the potentials, so far included, are $V(\phi)=V_0\phi^{\mu}e^{\beta \phi^\alpha}$ 
and $V(\phi)=V_0(\cosh(\alpha \phi) + \beta)$ with $\alpha$, $\mu$ and $\beta$ being free 
parameters, but the analysis can be easily extended to any other scalar field potential.  
Additional to the field component and the standard content of matter, the study
also incorporates the contribution from spatial curvature ($\Omega_k$), as it has been the 
focus in recent studies.
The analysis contains the most up-to-date datasets along with a nested sampler to produce 
posterior distributions along with the Bayesian evidence, that allows to perform a model selection.
In this work we constrain the parameter-space describing the two generic potentials, and amongst 
several combinations, we found that the best-fit to current datasets is given by a model
slightly favouring the quintessence 
field with potential
$V(\phi)=V_0\phi^\mu e^{\beta \phi}$ with $\beta=0.22\pm 1.56$, $\mu = -0.41\pm 1.90$, and slightly
negative curvature $\Omega_{k,0}=-0.0016\pm0.0018$, which presents deviations of $1.6\sigma$ from
the standard $\Lambda$CDM model. 
Even though this potential contains three extra parameters, the Bayesian evidence 
$\mathcal{B}_{\Lambda, \phi} =2.0$ is unable to distinguish this model compared to 
the $\Lambda$CDM with curvature ($\Omega_{k,0}=0.0013\pm0.0018$).
The potential that provides the minimal Bayesian evidence corresponds to 
$V(\phi)=V_0 \cosh(\alpha \phi)$ with $\alpha=-0.61\pm 1.36$.\\
\end{abstract}

\maketitle

\section{Introduction}

In the standard cosmological model, dark energy (DE) is considered as the source that drives 
the current accelerated expansion of the universe and also the dominant component, being around 70\% of 
the total matter-energy content.
The observational evidence and theoretical consistency of DE is well supported, 
for a review see \cite{Huterer:2017buf}. 
However, being either a material fluid or geometry, the physical mechanism behind DE 
is still a mystery. 
The simplest model for the DE is the cosmological constant added to the Einstein field equations. 
This idea with the addition of the cold dark matter component (CDM) are the foundations of the 
standard cosmological model or $\Lambda$CDM.
Despite being the simplest approach and even if the $\Lambda$CDM model fits well the current 
cosmological observations, it has been shown that a cosmological constant carries several 
issues of fundamental nature collectively known as ``cosmological constant problems'' 
\cite{lambda-p1, lambda-p2, lambda-p3, Zlatev:1998tr, lambda-p5, lambda-p6, lambda-p7}.   
This points out that the cosmological constant as DE needs a deeper study from fundamental 
physics and perhaps it is an approximation from a more complex model.

An alternative to the cosmological constant are the so called dynamical DE models, 
where the DE equation of state (EoS) has the form of a barotropic perfect fluid $p=w(t)\rho$.
In $\Lambda$CDM the EoS parameter $w$ of DE is a constant, $w_{\Lambda}=-1$, and hence implies a constant 
energy density, $\rho_{\Lambda}= const.$ with a constant negative pressure $p_{\Lambda}=-\rho_{\Lambda}$; 
in contrast, dynamical DE with varying $w(t)$ gives an evolving energy density, i.e, $\rho(t)$.
Furthermore, model independent techniques, based on observational data and on the minimal assumptions, 
are able to reconstruct the DE EoS and the results are in favour of a dynamical 
(time-dependent) EoS \cite{Vazquez:2012ce, Hee:2016nho, Zhao:2017cud, Wang:2018fng, Tamayo:2019gqj}.
The dynamical DE models provide alternatives to alleviate the cosmological constant problem amongst 
other important current conundrums in cosmology, like the recent $H_0$ tension
\cite{Yang:2018qmz, Pan:2019gop}.

Within this approach, scalar fields play a very important role and lead to particular dynamical DE models such as quintessence \cite{quint-1, quint-2, quint-3, quint-4, quint-5, quint-6, quint-7, Caldwell:2005tm, Tsujikawa:2013fta}, 
phantom DE \cite{phantom-1, phantom-2, phantom-3, phantom-4, Kujat:2006vj, Ludwick:2017tox}, 
$k$-essence \cite{k-1, k-2, k-3, k-4}, quintom DE \cite{qton-1, qton-2, qton-3, qton-4, Cai:2009zp}, 
among many others.
The main idea consists of introducing a scalar field whose associated energy density is able to 
mimic the cosmological constant at late times. 
The most popular scalar field DE models contain a single minimally coupled scalar field with 
a kinetic energy term, where the positive sign corresponds to quintessence and the negative for phantom, 
and a given scalar field potential. The energy density is the sum of the kinetic energy and the potential 
whereas their difference results in the effective pressure.
The potential of the scalar field supplies the required negative pressure to drive the accelerated
expansion of the Universe and its evolution; consequently the time evolution of the EoS depends crucially on the functional form of the potential.

In this work we focus on two scenarios: quintessence and phantom. 
Quintessence is a canonical scalar field minimally coupled to gravity \cite{Tsujikawa:2013fta}; 
it is considered as the simplest scenario with no theoretical problems such as the appearance of 
ghosts or Laplacian instabilities, it describes a time-evolving DE which alleviates the 
cosmological constant problem. 
For instance, the so-called 
``coincidence problem'', namely why the dark matter and dark energy density 
happen to be of the same order today \cite{coinc-p}.
Phantom, on the other hand, is a non-canonical scalar field in which the kinetic energy is 
negative \cite{1,2,3,4,5}, despite the ``wrong" sign of the kinetic term, phantom models are 
able to resolve the $H_0$ tension \cite{1}; meanwhile, in \cite{4} the viability of phantom fields
is demonstrated, even when quantum effects are taken into account. 
This is a very important result since one of the strongest criticisms to the phantom 
fields is the fact that, since their kinetic energy density is negative, at first sight, 
one cannot construct a healthy quantum theory of this field. 
But in \cite{4}, up to first order perturbations in a FLRW background, the authors computed 
the expectation value of the field's kinetic energy, demonstrating that there is a region in 
the parameter space where it is not a negative quantity.

As mentioned before, the functional form of the potential $V(\phi)$ determines the time evolution 
of the scalar field and, consequently, also that of $w(z)$. 
Since we don't have yet a derivation of $V(\phi)$ from cosmological principles, the common 
approach is to propose functional forms of the potential inspired from particle physics, 
mostly in a phenomenological manner, to see how it fits to the data and solves some cosmological challenges.
In the literature there is a plethora of proposals for scalar field potential for 
late time acceleration \cite{Ratra:1987rm, Ferreira:1997hj, Zlatev:1998tr, Brax:1999gp, Caldwell:2005tm, Scherrer:2008be, Dutta:2008qn, Dutta:2009dr, Chang:2016aex}. 
Given a potential, one can always constrain its parameters using the current available set of 
cosmological data. A step further is to test a collection of potentials and compare its 
statistical viability in terms of current observations
\cite{Avsajanishvili:2017zoj,Lonappan:2017lzt,Durrive:2018quo,Roy:2018nce,Tosone:2018qei,Yang:2018xah}. 

Following this aim, instead of studying a particular scalar field DE potential or a group of them, 
in this work we propose two generic forms of potentials: $V(\phi)=V_0\phi^{\mu}e^{\beta \phi^\alpha}$ and 
$V(\phi)=V_0(\cosh(\alpha \phi) + \beta)$ ($\alpha$, $\mu$ and $\beta$ being free parameters). 
The advantage of our proposal is that it encompasses several potentials proposed in 
the literature; for specific combinations of values of the free parameters, these generic 
potentials reproduce known scalar field potentials, for both quintessence and phantom 
cases depending on a switch parameter $\epsilon$, which will be shown in the following section.
In addition, we also use the spatial curvature density parameter $\Omega_{k,0}$ 
as a free parameter. 
   
The paper is organised as follows.
In Sec. \ref{Scalar Field Equations}, we set up the mathematical background of minimally coupled scalar fields in cosmology and its description as a dynamical system, we then introduce 
the switch parameter $\epsilon$ in order to have a joint description of quintessence and phantom.
In Sec. \ref{Initial conditions}, we estimate the initial conditions of the variables of the 
dynamical system to be solved.
We present the generic potentials and their links with other particular 
potentials in Sec. \ref{Potentials}, and in Sec. \ref{Code and Datasets}, we introduce the code 
used throughout the analysis and the observational datasets included. In Sec. \ref{Results}, we show the 
posteriors of the Bayesian analysis, the constraints of the model parameters in 
several particular cases and a model comparison through the Bayesian evidence. 
Finally in Sec. \ref{Conclusions} a summary of our results and an outlook for future research is given. 

Along this work we use the units $\hbar=c=8\pi G = 1$.

\section{Scalar Field Equations}\label{Scalar Field Equations}

The action of a cosmological model including a minimally coupled single scalar field $\phi$, 
either quintessence or phantom, is
\begin{equation}
    S = \int d^4x\sqrt{-g}\left[\frac{R}{2} +\epsilon \frac12 \partial^\nu \phi \partial_\nu \phi  -V(\phi)  + \mathcal{L}_M\right], \nonumber
\end{equation}
where $g$ is the determinant of the metric $g_{\mu \nu}$, $R$ is the Ricci scalar and  
the term $\mathcal{L}_M$ accounts 
for the other matter components of the universe (namely dark matter, baryons, radiation, etc.).
Here, to distinguish the type of field, we have introduced the switch parameter
\begin{equation} \label{eq:epsilon}
	\epsilon =\left\{
        	 \begin{array}{ll}
         	      +1  \qquad {\rm quintessence},  \\
         	       \phantom{}\\
         	       -1 \qquad {\rm phantom}.
         	       \end{array}
              \right.
\end{equation}

\noindent
Considering a FLRW universe, 
the Friedmann equations are thus
\begin{eqnarray}
    H^2+\frac{k}{a^2} &=& \frac{1}{3}(\rho_\phi +\rho_M), \label{eq:Friedman1}\\
    \dot{H}-\frac{k}{a^2} &=& -\frac{1}{2}(\rho_\phi +p_\phi +\rho_M+p_M), \label{eq:Friedman2}
\end{eqnarray}
where $H=\dot a/a$, $a$ is the scale factor, over-dot describes time derivative 
$\dot x =dx/dt,$ and $k$ the intrinsic curvature. 
The standard matter components, $\rho_M = \sum \rho_i$, are assumed as perfect fluids 
and have a barotropic EoS of the form $p_i=w_i\rho_i$, hence the usual energy 
conservation equation for each one reads as
\begin{eqnarray}
    \dot{\rho}_i+ 3H (1+w_i) \rho_i = 0.
\end{eqnarray}
In the case of pressureless matter we have $w_i=0$, whereas for the relativistic 
particles $w_i=1/3$; similarly, the curvature can be considered as an effective perfect fluid
with equation of state $w_i=-1/3$, and move it to the right-hand-side on the expressions
(\ref{eq:Friedman1}) and (\ref{eq:Friedman2}). 
For the scalar field, the effective energy density and pressure are given by,
\begin{eqnarray}\label{rho p}
    \rho_\phi = \epsilon \frac{1}{2}\dot{\phi}^2 +V(\phi), \label{quintom energy density}\\
    p_\phi = \epsilon \frac{1}{2} \dot{\phi}^2 -V(\phi). \label{quintom pressure}
\end{eqnarray}
The associated EoS of the scalar field then becomes
\begin{equation}
	w_\phi = \frac{\epsilon\dot \phi^2 - 2V(\phi)}
	{\epsilon \dot \phi^2 + 2V(\phi)} \, ,
\end{equation}
whose value can be determined from the evolution of the field itself that
satisfies the Klein-Gordon equation 
\begin{equation}\label{eq:KGs}
    \ddot\phi + 3H\dot\phi +\epsilon V_{,\phi} = 0,
\end{equation}
where $x_{,\phi} = dx/d\phi$.
Following previous papers 
\cite{Lazkoz:2007mx, Scherrer:2007pu, Matos:2008ag, Scherrer:2008be, Gonzalez:2008wa}, 
the equations of motion are written in the form of a dynamical system 
by introducing the variables:
\begin{eqnarray} \label{variables}
    \Omega_i &=&\frac{\rho_i}{3H^2}, \qquad 
    \Omega_\phi =\frac{ \rho_\phi}{3H^2}, \nonumber \\
    \lambda &=& -\epsilon \frac{V_{,\phi}}{V}, \quad \Gamma = V\frac{V_{,\phi \phi}}{V_{,\phi}^2}.
\end{eqnarray}
The minus sign in the definition of $\lambda$ corresponds to $\dot \phi>0$ 
($V_{,\phi} <0$) (for the opposite there would be a plus sign) such that $\lambda$ remains
positive; for further details see \cite{Chiba:2005tj}.
This convention allows to write $\epsilon$ in terms of the potential and its derivative
through the \textit{sign} function
\begin{equation} \label{eq:eps_V}
    \epsilon = {\rm sign}(-\partial_\phi \ln V(\phi) ),
\end{equation}
and we can then introduce
\begin{equation} 
	\teps \equiv  -\partial_\phi \ln V(\phi) =\left\{
        	 \begin{array}{ll}
         	      >0  \qquad {\rm quintessence},  \\
         	       \phantom{}\\
         	       <0 \qquad {\rm phantom},
         	       \end{array}
              \right.
\end{equation}
such that 
\begin{equation}
\lambda=\teps \epsilon =\teps~ \siteps.    
\end{equation}
Thus, both types of fields are identified in a single function $\teps$.
Important to stress out, there is no crossing of the phantom-divide-line, but this parameter allows to put both models within the same analysis. This approach will also be useful when combining both
types of fields, i.e. quintom models. 
\\

Therefore, the dynamical system to solve turns out to be
\begin{eqnarray} \label{eq:dynamical}
\Omega_i' &=& -3\Omega_i \left( 1 + w_i +  \frac{2}{3}\frac{\dot H}{H^2}\right),  \\
\Omega_\phi' &=& -3\Omega_\phi \left( 1 + w_\phi +  \frac{2}{3}\frac{\dot H}{H^2}\right), 
\nonumber \\
w_\phi' &=& -(1-w_\phi) \left(3(1+w_\phi) -\teps  \sqrt{3\Omega_\phi(1+w_\phi)\siteps } \right), \nonumber \\
\teps' &=& -\teps^2 (\Gamma -1 ) \sqrt{3\Omega_\phi(1+w_\phi)\siteps  }, \nonumber 
\end{eqnarray}
\noindent
where prime indicates derivative with respect to the e-fold number $x'\equiv dx/d\ln a$. 
The last term of the first two equations of \eqref{eq:dynamical} is written as 
\begin{equation}\label{eq:Hub}
    -\frac{\dot H }{H^2} = q+1=\frac{3}{2}\left [\sum_i w_i \Omega_i + w_\phi\Omega_\phi +1 \right],
\end{equation}
where $q$ is the deceleration parameter, 
and we have made used of the Friedmann equation 
\begin{equation} \label{eq:Fried}
    1 = \sum_i \Omega_i + \Omega_\phi.
\end{equation}
The associated equation of state of the scalar field can be described as a 
deviation of the cosmological constant, 
\begin{equation}\label{eq:eos}
       w_\phi =-1+ \siteps \gamma,
\end{equation}
with a positive function $\gamma$. Moreover, the value of $\teps$ will identify the type of 
field to be considered, either quintessence ($+1$) or phantom ($-1$), and hence the terms 
within the square root [$(1+w_\phi)\siteps$] in the dynamical system  
will remain positive.

Notice that not all  equations in the dynamical system (\ref{eq:dynamical}) are linearly independent, that is, either one of the components of the matter fluids $\Omega_i$, or the scalar field component $\Omega_\phi$, can be written as a linear combination of the remaining ones. 
Hence, the dimension of the phase space equals $d=3+(N-1)=N+2$, where $N$ is the number 
of matter fields. As in our case, since we choose only two additional components 
(matter and curvature),  
then, the derivative of the curvature term, corresponding to $\Omega_k= \rho_k/ (3H^2)$, is 
given by the first expression in (\ref{eq:dynamical}) 
\begin{equation}
\Omega_k' = -3\Omega_k \left (1 -\frac{1}{3} + \frac{2}{3} \frac{\dot H}{H^2} \right ).    
\end{equation}
Also, in (\ref{eq:Hub}) both the matter and curvature contribution will be given by 
$\sum_i w_i \Omega_i=-\Omega_k/3$, 
and due to the Friedmann constraint (\ref{eq:Fried}) we can eliminate $\Omega_m$ 
in the dynamical system. 
Therefore the independent phase-space variables  are, $\Omega_\phi$, $w_\phi$,  
$\teps$ and  $\Omega_k$, i.e., the phase space is 4-dimensional.
The dimension of the phase space can be further reduced if an exponential potential is chosen since, in these cases, $\teps$ is a constant, as we shall see in the next section.

\section{Initial conditions}\label{Initial conditions}

Even though the system can be extended for any amount of components$-i$, 
for the sake of this work we restrict to a universe made of a scalar field, dust 
(dark matter + baryons) 
and curvature, and considering a smooth transition to the radiation domination 
epoch (photons and three neutrino species, $N_{\rm eff}=3.046$, with minimum allowed 
mass $\sum m_\nu=0.06$ eV). 

Some papers pointed out the sensitivity of the initial conditions in order to get accurate results \cite{Kujat:2006vj}. 
There have been several approaches for the conditions, for instance by assuming a general 
potential $V(\phi)=V_0 f(\phi)$, in \cite{Avsajanishvili:2017zoj} used $V_0$, $\phi_0$, $\dot \phi_0$ as free parameters for the initial conditions, additional to the parameters that describe the form of the potential.

In our case, the initial conditions can be set up right into the matter domination epoch ($a_{\rm ini}\sim 10^{-3}$) and we are considering the initial conditions that realize thawing behavior (the field is initially frozen at the flat part of the potential due to large Hubble friction and behaves like cosmological constant $w=-1$ \cite{Scherrer:2007pu}), then we are able to choose the initial EoS of the scalar field ($w_{\phi,{\rm ini}}$) very close to the cosmological constant. 
Therefore, from Eqn. (\ref{eq:eos}) we keep fixed $\gamma_{\rm ini}=10^{-4}$, for either sign of $\teps$. 

The initial value for the field density parameter, $\Omega_{\phi,{\rm ini}}$, can be taken as a free parameter. However, we tested this process and adding an extra parameter will reduce the acceptance rate in the analysis and therefore increase the computation time considerably.
Thus, to enhance the process and minimise the computation time, 
$\Omega_{\phi,{\rm ini}}$ is thus selected such that its present value is $\Omega_{\phi,0}=1-\sum_i \Omega_{i,0}$.
This can easily be achieved with a shooting method. 

The parameter $\teps_{\rm ini}$ will decide the type of field in place, and  can be 
 either chosen
as a sampling parameter or instead of taking the initial value of the 
field $\phi_{\rm ini}$. 
We performed a Bayesian analysis by using both parameters and the results, as expected, will produce similar constraints on the base parameters.
However the selection of using $\phi^{-1}_{\rm ini}$ provides a  
slightly better fit and less correlated constraints over the selection of $\teps_{\rm ini}$.

\begin{figure}[t!]
\includegraphics[trim = 0mm  0mm 0mm 1mm, clip, width=8cm, height=6.cm]{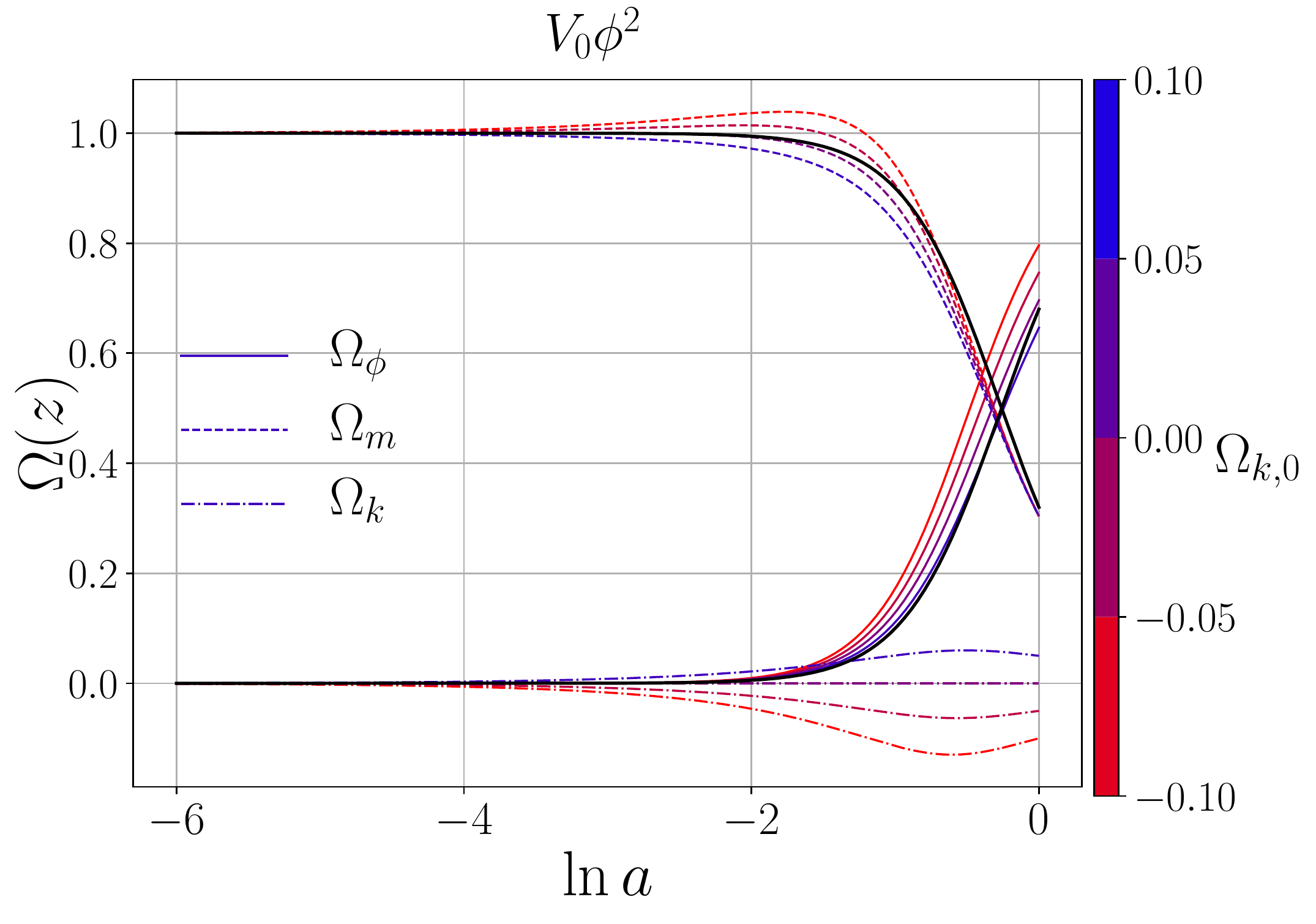}
\caption[Quintom2]{Density parameters for a Universe where DE is described by a quintessence field with potential $V(\phi)=V_0\phi^2$, dust described by dark matter $+$ dust ($\Omega_m$) and a curvature component ($\Omega_k$) coloured coded to span the range of $\Omega_{k,0}=[-0.1, 0.1]$. Black solid lines describe the flat-$\Lambda$CDM model by using the current Planck values.}
\label{Fig:density}
\end{figure}

To illustrate the whole process, Figure \ref{Fig:density} displays the density parameters for a Universe where DE is described by a quintessence field ($\teps_{\rm ini}>0$) with potential $V(\phi)=V_0\phi^2$, dust describes the dark matter + baryons ($\Omega_m$) and a
curvature component ($\Omega_k$) is coloured coded to span the range $\Omega_{k,0}=[-0.1, 0.1]$. 
The initial conditions are fixed right into the matter domination epoch.
Black solid lines describe the flat-$\Lambda$CDM model by using the current Planck values. 

 \section{Potentials}\label{Potentials}

The main aim of this work is to study a
general potential, for either quintessence or phantom, and 
compares its observables with current cosmological observations;
then to include a python module into the SimpleMC code \cite{SimpleMC, Aubourg:2014yra}. 
As a proof of the concept, here we focus on two general classes of potentials that comprise a wide variety of fundamental models of particular importance for cosmology. First let us explore the three-parametric class of potentials: (a)  $V(\phi)= V_0 \phi^\mu e^{\beta \phi^\alpha}$ and then the two-parametric class: (b) $V(\phi)= V_0(\cosh(\alpha \phi) + \beta$).

\begin{figure*}[ht!]
\begin{center}
\includegraphics[trim = 0mm  0mm 0mm 1mm, clip, width=5.5cm, height=6.cm]{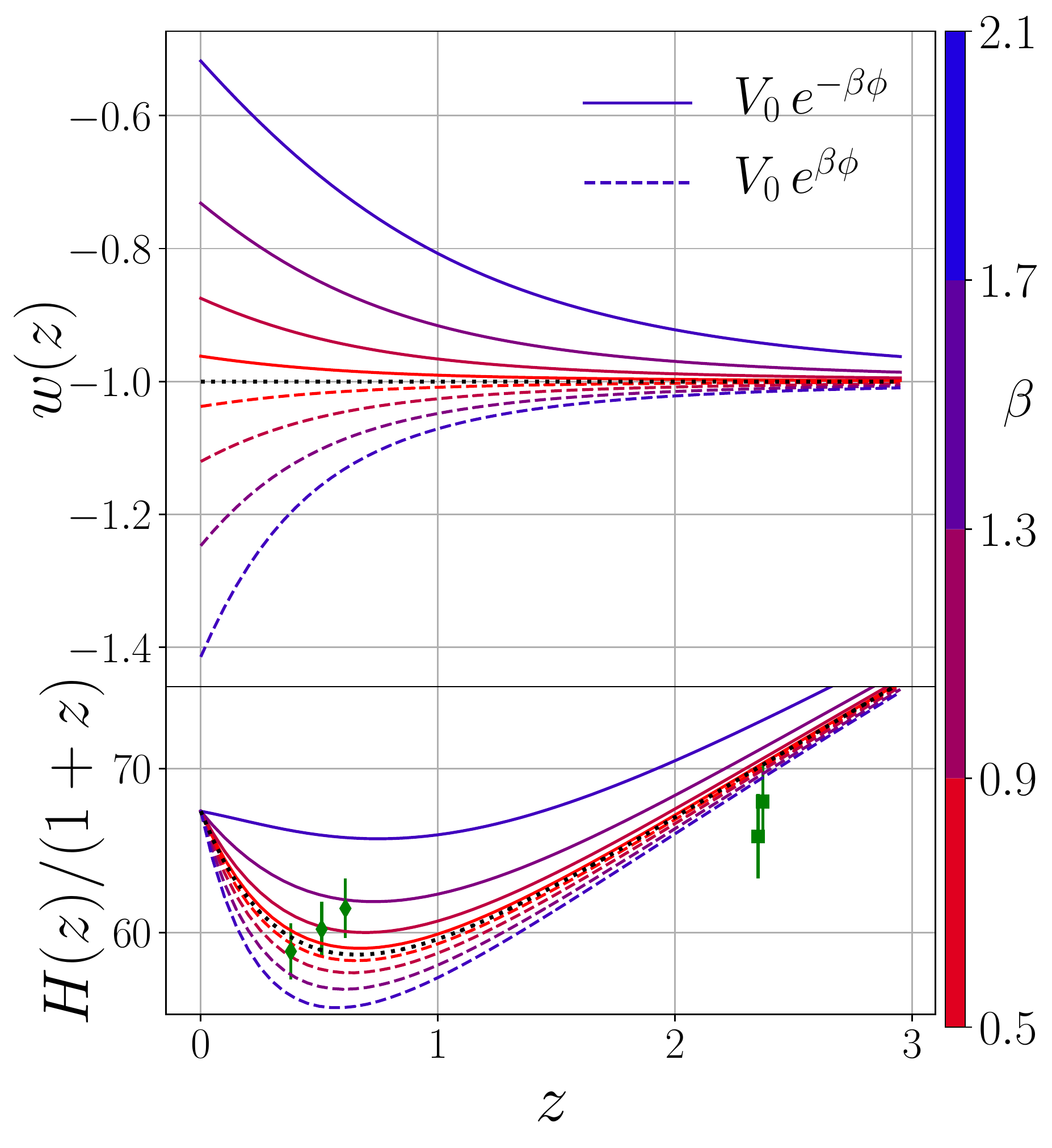}
\includegraphics[trim = 0mm  0mm 0mm 1mm, clip, width=5.5cm, height=6.cm]{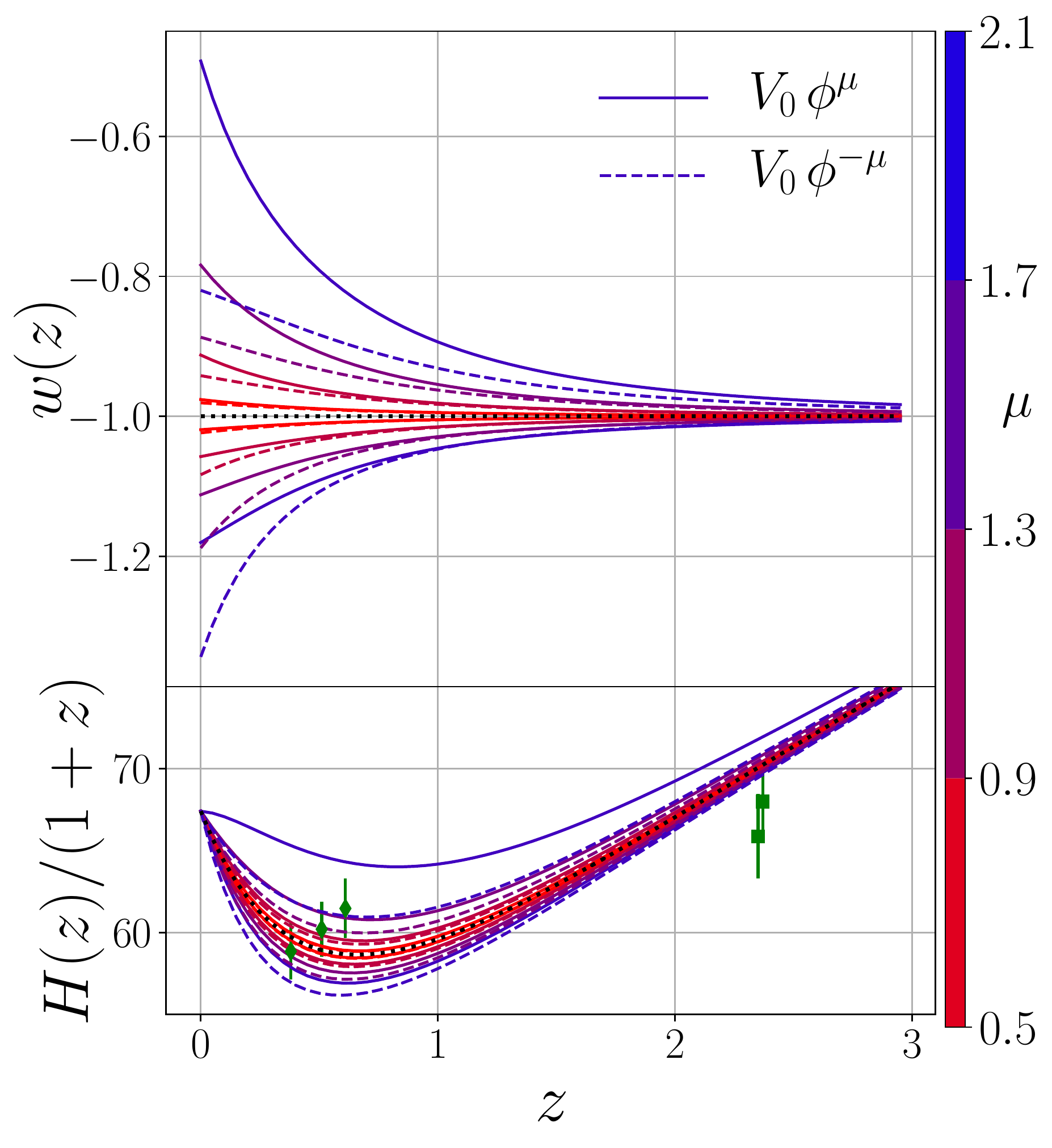} 
\includegraphics[trim = 0mm  0mm 0mm 1mm, clip, width=5.5cm, height=6.cm]{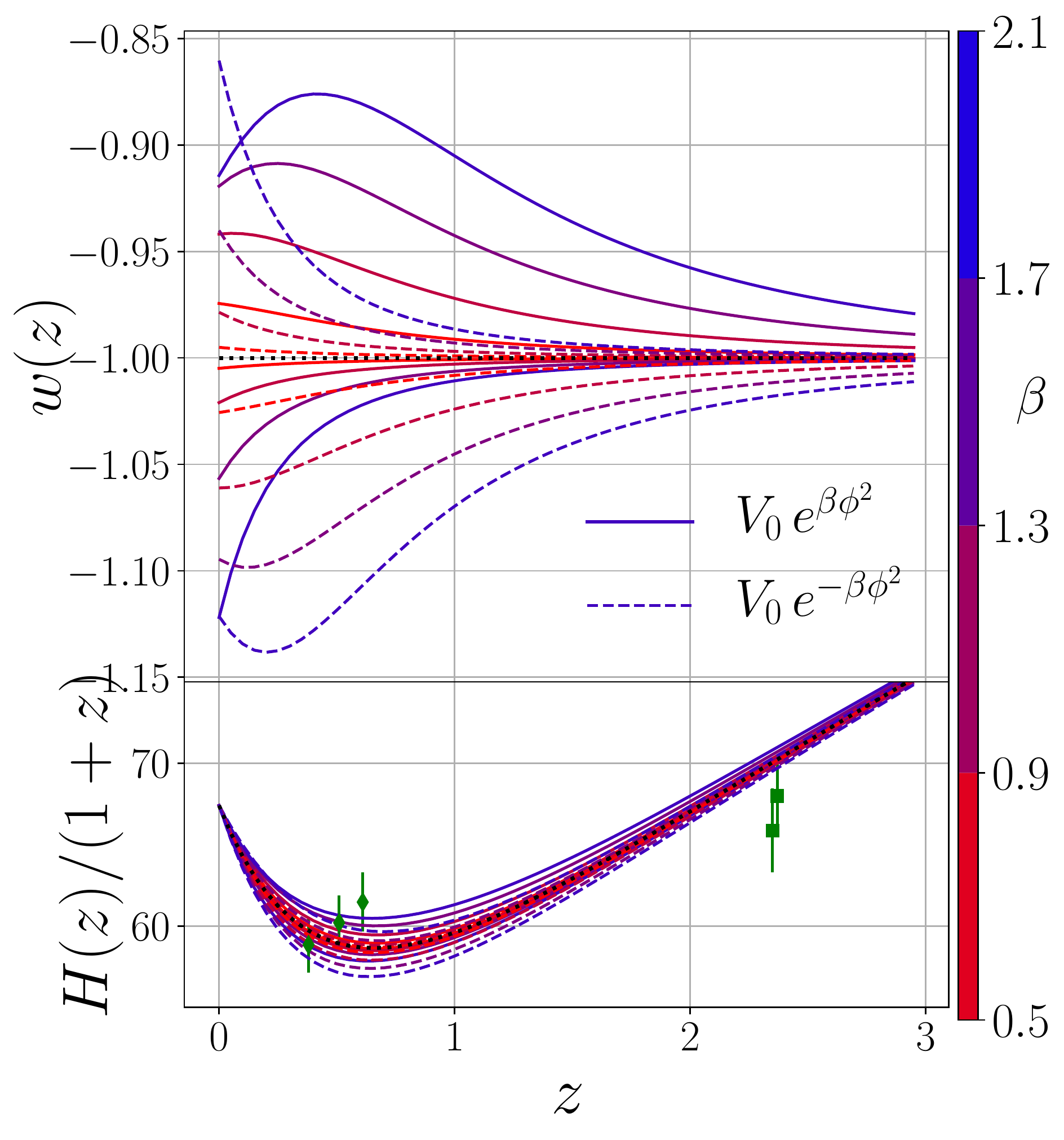}  
\includegraphics[trim = 0mm  0mm 0mm 1mm, clip, width=5.5cm, height=6.cm]{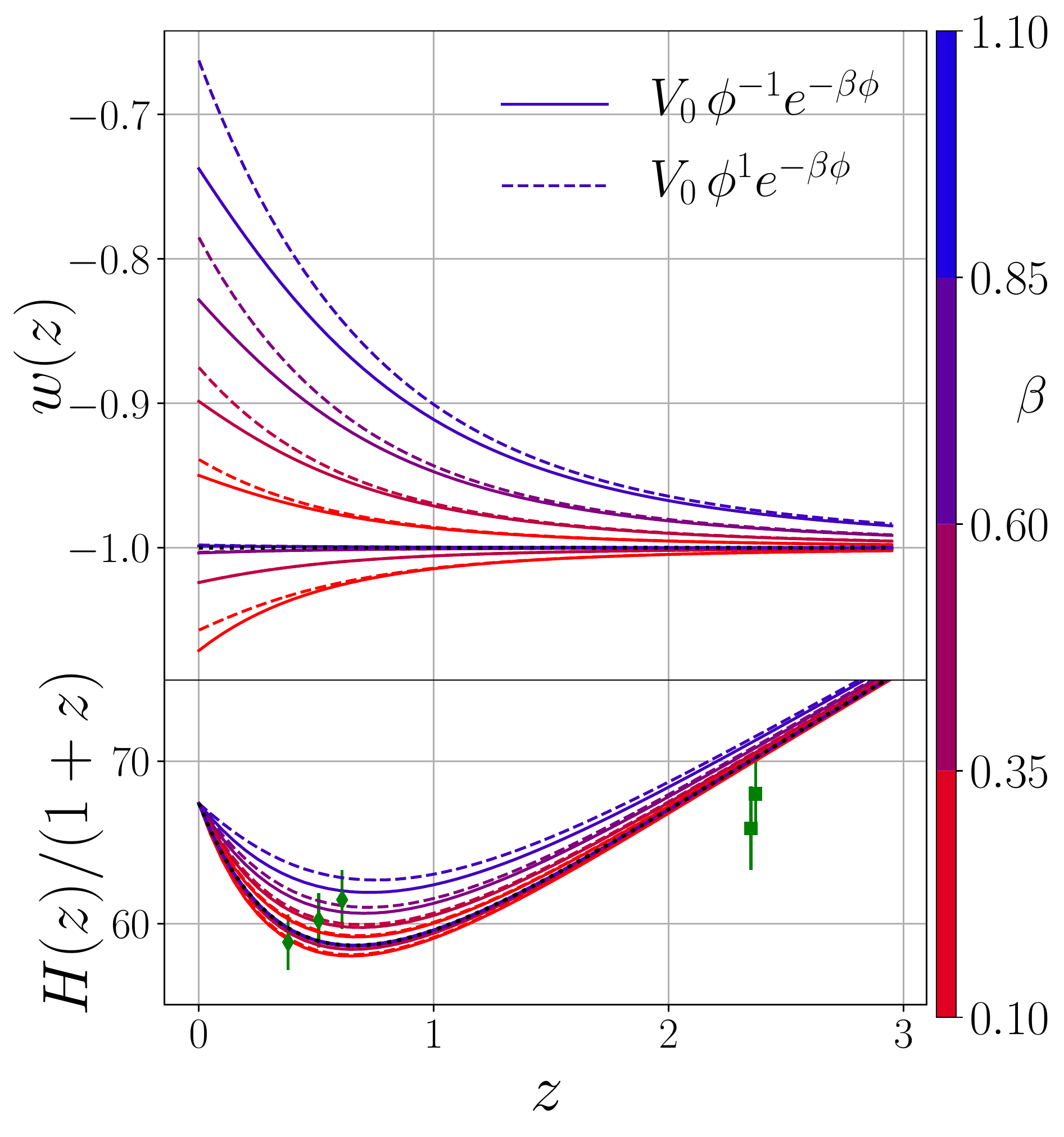} 
\includegraphics[trim = 0mm  0mm 0mm 1mm, clip, width=5.5cm, height=6.cm]{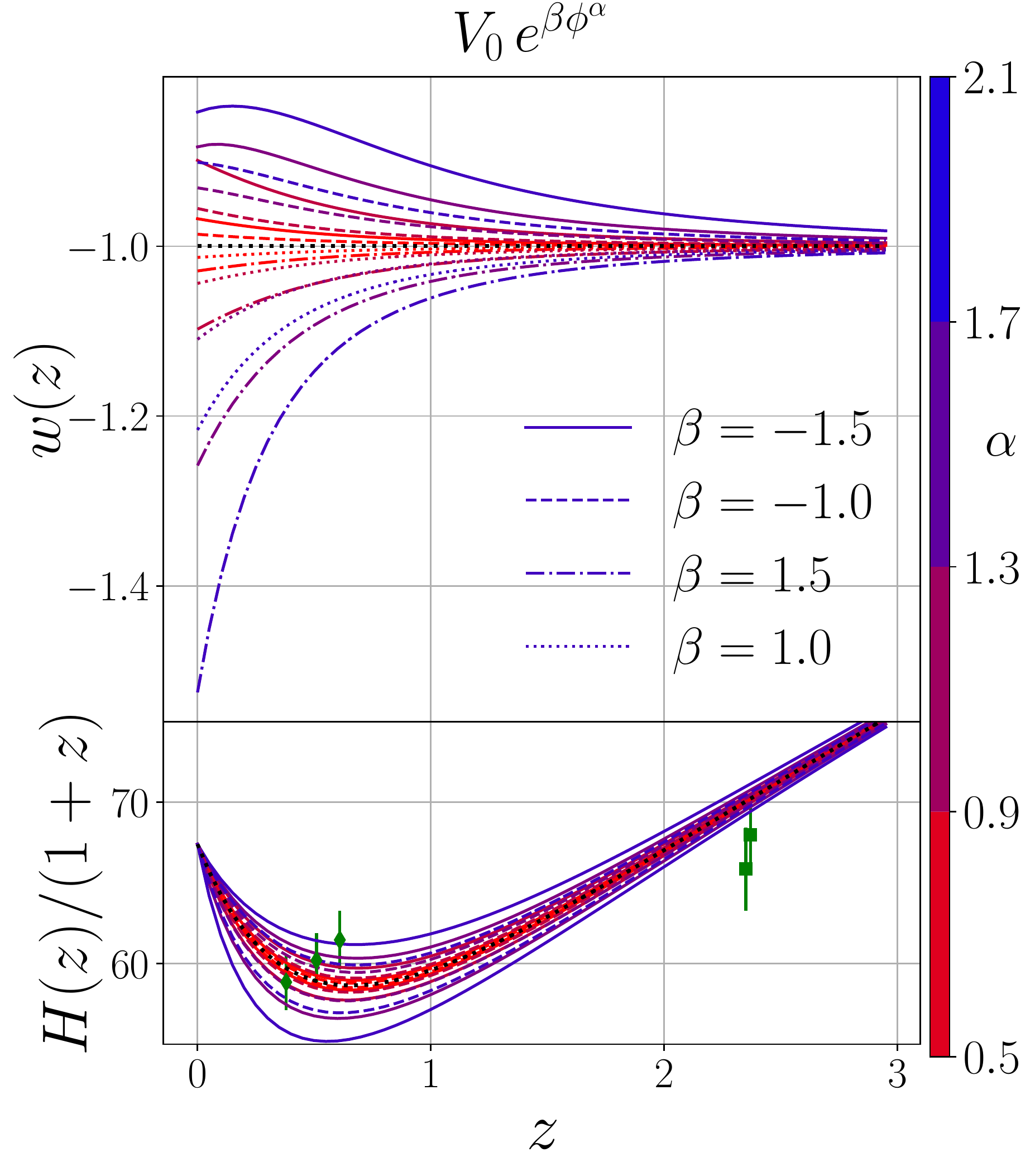} 
\includegraphics[trim = 0mm  0mm 0mm 1mm, clip, width=5.5cm, height=6.cm]{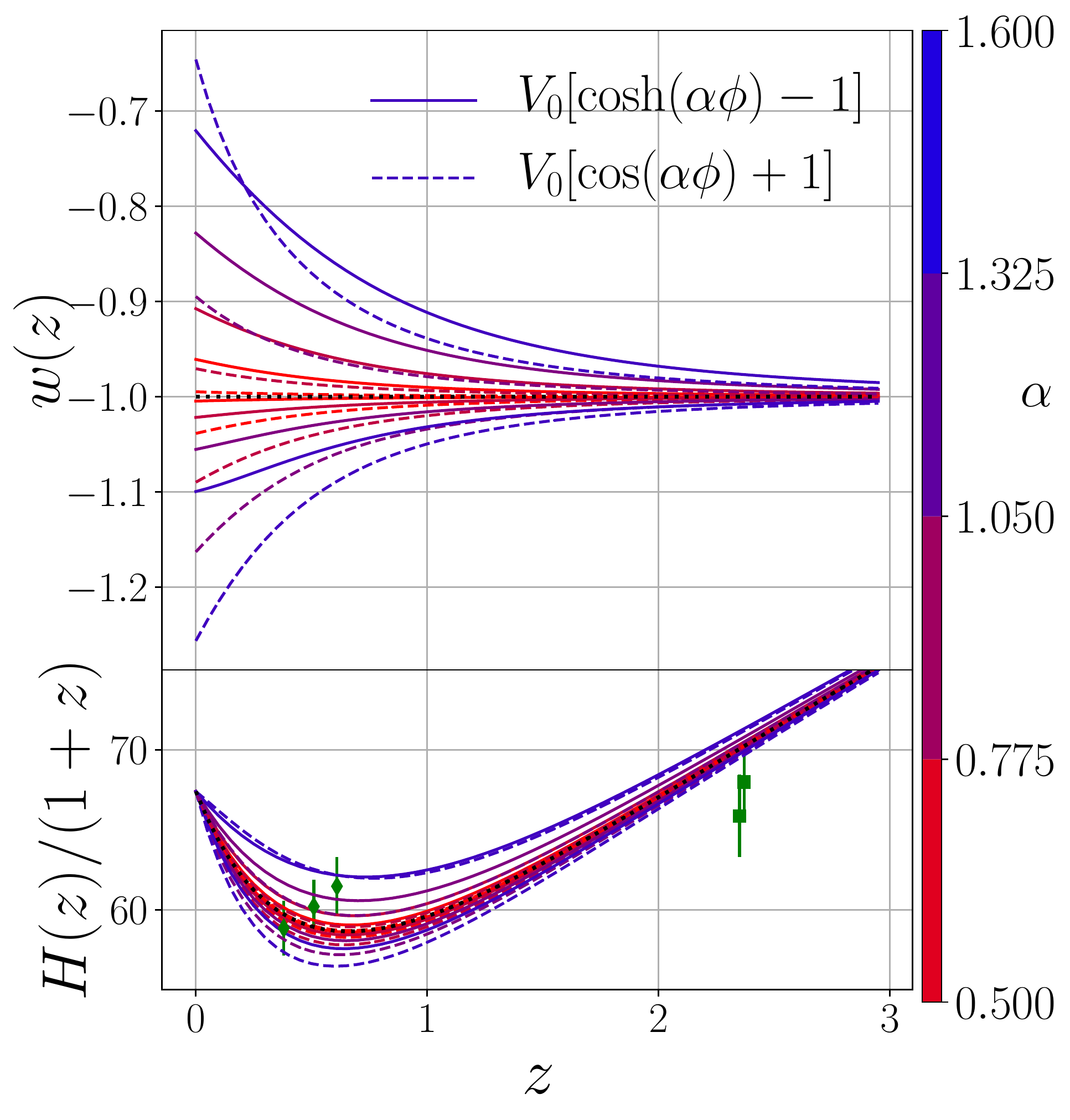} 
\end{center}
\caption[Quintom2]{Evolution of the DE EoS parameter $w(z)$ for selected values of the parameters of the generic potentials (both quintessence and phantom) and $H(z)/(1+z)$,  along with data from BAO Galaxy consensus
and Ly-$\alpha$ DR14.}
\label{Fig:model_QP}
\end{figure*}

The generic potential (a) under particular choices of the parameters $\mu$, $\beta$ and $\alpha$ boils down into  potentials already studied in the literature. 
Here, $\alpha, \lambda, \chi, \gamma$ being positive constant values, we have:
\[(a) \quad V(\phi)= V_0\phi^\mu e^{\beta \phi^\alpha}. \]
For quintessence:
\begin{itemize}
    \item $\mu=-\alpha$, $\beta=0$: 
        $V(\phi) = V_0 M^2_{pl} \phi^{-\alpha}$, \cite{Ratra:1987rm}.
\item $\mu=0$, $\beta=-\lambda/M_{pl}$, $\alpha=1$: 

$V(\phi) = V_0 \exp[-\lambda \phi/M_{pl}]$, \cite{Ferreira:1997hj}.
\item $\mu=-\chi$, $\beta = \gamma/M^2_{pl}$, $\alpha=2$: 

$V(\phi) = V_0 \phi^{-\chi} \exp[\gamma \phi^2/M^2_{pl}]$, 
\cite{Brax:1999gp}.
\item $\mu=0$, $\beta = M_{pl}$, $\alpha=-1$: 

$V(\phi) = V_0 \exp[M_{pl}/\phi]$, \cite{Caldwell:2005tm}.    
\end{itemize}
For phantom:
\begin{itemize}
    \item  $\mu=6$, $\beta=0$: $V(\phi) = V_0 \phi^6$, \cite{Scherrer:2008be}.
    \item $\mu=-2$, $\beta=0$: $V(\phi) = V_0 \phi^{-2}$, \cite{Scherrer:2008be}.
    \item $\mu=0$, $\alpha=1$: $V(\phi) = V_0 e^{\beta \phi}$, \cite{Scherrer:2008be}.
    \item $\mu=2$ $\beta=0$: $V(\phi) = V_0 \phi^2$, \cite{Dutta:2009dr}.
\end{itemize}
In \cite{Avsajanishvili:2017zoj} the authors make a joint study about several of 
these potentials.

Similarly for the second potential: 
\[(b) \quad V(\phi)= V_0( \cosh(\alpha \phi) + \beta). \]
\begin{itemize}
\item $\beta=1$, $\alpha=i/f$: $V(\phi) = V_0 (\cos(\phi/f) + 1)$, \cite{Dutta:2008qn}.
\item $\beta=-1$: $V(\phi) = V_0 (\cosh(\alpha \phi) -1)$, \cite{Sahni:1999qe}.
\item $\beta=0$: $V(\phi) = V_0 \cosh(\alpha \phi)$, \cite{UrenaLopez:2000aj}.
\end{itemize}
In general, several of these potentials have been also studied in the context of dark matter and Inflation, for instance $V_0\phi^2$ \cite{Matos:2008ag}, $V_0\lambda\phi^4$ \cite{Padilla:2019fju}, $V_0(\cosh(\lambda \phi)-1)$ \cite{Matos:2009hf}, $V_0(a\cos(\phi/f)+1)$ \cite{Ross:2016hyb}, $V_0e^{\phi^2}(\phi-\phi_0)^2$ \cite{German:2019aoj}.

Notice that once we specify the scalar field potential, we are able to compute $\teps$ and $\Gamma$ (given in the expressions (\ref{variables})) and hence the dynamical system (\ref{eq:dynamical}) 
will be closed.
For the potential $V(\phi)= V_0\phi^\mu e^{\beta \phi^\alpha}$ we have the $\teps$ and $\Gamma$  
functions are respectively
\begin{eqnarray}
    - \teps &=& \mu \phi^{-1} + \alpha \beta \phi^{\alpha-1}, \nonumber \\
    \Gamma &=& 1 + \frac{1}{\teps^2}\left[\alpha(\alpha-1)\beta \phi^{\alpha-2} - \mu \phi^{-2} \right].
\end{eqnarray}
and similarly for $ V(\phi)= V_0(\cosh(\alpha \phi) + \beta)$ we have
\begin{eqnarray}
    - \teps &=& \frac{\alpha\sinh(\alpha \phi)}{\cosh(\alpha \phi)+ \beta}, \nonumber \\
    \Gamma &=&  -\frac{\alpha}{\teps}\coth(\alpha \phi).
\end{eqnarray}
\noindent

With these quantities in mind, the free parameters can be sampled over and find out their
constraints. 
To cover several of the aforementioned parameters, and furthermore, Table \ref{TableI} 
shows some of these particular cases. 
Figure \ref{Fig:model_QP} displays, for the potentials we focus on, the general behaviour 
of its equation of state along with the $H(z)/(1+z)$ function. Green error bars correspond to the
 galaxy and Ly-$\alpha$ BAO, as we shall explain below.

\begin{table}[ht!]
\captionsetup{justification=raggedright,singlelinecheck=false,font=footnotesize}
\footnotesize
\scalebox{0.95}{%
\begin{tabular}{cccccc} 
\cline{1-6}\noalign{\smallskip}
 \vspace{0.15cm}
$\beta$ \quad & $\mu$ \quad& $\alpha$  \quad& Model & $\teps$  & $\Gamma$  \\
\hline
\vspace{0.15cm}
&&&&&\\
\vspace{0.15cm}
$\beta$ \quad&  $0$ \quad & $1$ \quad& $e^{\beta \phi}$  & $- \beta$ &  $1$  \\
\vspace{0.15cm}
0 &  $\mu$ & $-$& $\phi^\mu$  & $- \mu \phi^{-1}$ &  $1- \cfrac{1}{\mu}$  \\
\vspace{0.15cm}
$\beta$ &  $0$ & $2$& $e^{\beta \phi^2}$  & $-2 \beta \phi$ &  $1+\cfrac{2\beta}{\teps^2}$  \\
\vspace{0.15cm}
$\beta$ &  $\mu$ & $1$& $\phi^\mu e^{\beta \phi}$ &$- (\mu\phi^{-1} +\beta)$& 
    $1-\cfrac{1}{\mu}\left( 1+ \cfrac{\beta}{\teps} \right)^2$   \\
\vspace{0.15cm}
$\beta$ &  $0$ & $\alpha$& $e^{\beta \phi^\alpha}$  & $- \alpha \beta \phi^{\alpha-1}$ &  
        $1 +\cfrac{(\alpha-1)}{\alpha \beta}\left(-\cfrac{\teps}{\alpha\beta} \right)^{\frac{\alpha}{1-\alpha}}$  \\
\vspace{0.15cm}
$\beta$ &  2 & 2 &  $\phi^2e^{\beta \phi^2}$ & $-2 (\phi^{-1} + \beta \phi)$ & $\cfrac{3}{4} +\cfrac{4\beta}{\teps^2} \mp\cfrac{\sqrt{\teps^2-16\beta}}{4\teps}$ \\
\hline
\vspace{0.3cm}
0 &  - & $\alpha$& $\cosh(\alpha \phi)$  & $-\alpha \tanh(\alpha \phi)$ &  $ \cfrac{\teps^2}{\alpha^2}$  \\
\vspace{0.15cm}
-1 &  - &  $\alpha$& $\cosh(\alpha \phi) - 1$ & $-\alpha\coth (\cfrac{\alpha \phi}{2})$  & $\cfrac12 \left(1+\cfrac{\alpha^2}{\teps^2}\right)$ \\
\vspace{0.15cm}
1 &  - & $i\alpha$& $\cos(\alpha \phi) + 1$  & $\alpha \tan(\cfrac{\alpha \phi}{2})$ 
    &  $\cfrac12 \left(1-\cfrac{\alpha^2}{\teps^2}\right)$ \\
\hline
\hline
\end{tabular}}
\caption{Selected parameters and the model they described, along with 
 the auxiliary variables $\teps$ and $\Gamma$. }
\label{TableI}
\end{table}

\section{Code and Datasets}\label{Code and Datasets}

To explore the parameter space and impose constraints on the free parameters, 
we use the new version of the SimpleMC code \cite{SimpleMC, Aubourg:2014yra}. 
The code already contains several samplers for a proper exploration of the parameter-space, 
but in particular we use a modified version of the nested sampler Dynesty \cite{Speagle:2020, Skilling:2004, Skilling:2006} 
that allows to explore complex posterior distributions and to 
compute the Bayesian evidence, which is used to 
perform a model comparison through the Jeffreys scale \cite{Vazquez:2011xa}.
The Bayesian Evidence has been used in several papers to compare DE models, parametrizations for the DE EoS, inflationary models and to perform reconstructions of cosmological functions, amongst many other  applications, see for instance \cite{Vazquez:2011xa, Vazquez:2012ux, Vazquez:2012ce, Hee:2016nho}. 
For a comprehensive review of the parameter inference in cosmology, see Ref. \cite{Padilla:2019mgi}.
Throughout this analysis we use  the recent BAO datasets from Ly-$\alpha$ DR14, BAO-Galaxy consensus, MGS and 6dFGS as presented in \cite{Alam:2016hwk, Blomqvist:2019rah, Ata:2017dya, Agathe:2019vsu, Beutler2011, Anderson:2013zyy},  a collection of currently available cosmic Chronometers that provides measurements of the Hubble function (see \cite{Gomez-Valent:2018hwc} and references therein), a compressed version of the Pantheon dataset which speeds up the process without loosing accuracy \cite{Scolnic:2017caz}, and a compressed version of Planck-15 information (treated as a BAO experiment located at redshift $z=1090$, \cite{Aubourg:2014yra}) 
to improve constraints and break degeneracies.

The flat priors of the base parameters used throughout the analysis correspond to $\Omega_{m,0}=[0.05,0.5]$ 
 for the matter density parameter today, $\Omega_{b,0}h^2$=[0.02, 0.025] for the physical 
baryon density parameter, $h=[0.4, 1.0]$  for the reduced Hubble constant and $\Omega_{k,0}=[-0.03, 0.03]$ 
for the effective curvature density parameter today. 
Whereas to select the priors that described the potential parameters we based upon 
Figure \ref{Fig:model_QP} which displays general behaviours for the functions $w(z)$ and $H(z)/(1+z)$ by varying 
some of these parameters. Hence, the flat priors we selected for the potential parameters are
$\mu =[-6,6]$, $\beta=[-3, 3]$, $\alpha=[-3,3]$, $\phi_{\rm ini}^{-1}=[-3,3]$.

\section{Results}\label{Results}

\begin{figure*}[ht!]
\begin{center}
\includegraphics[trim = 0mm  0mm 0mm 0mm, clip, width=5.5cm, height=4.5cm]{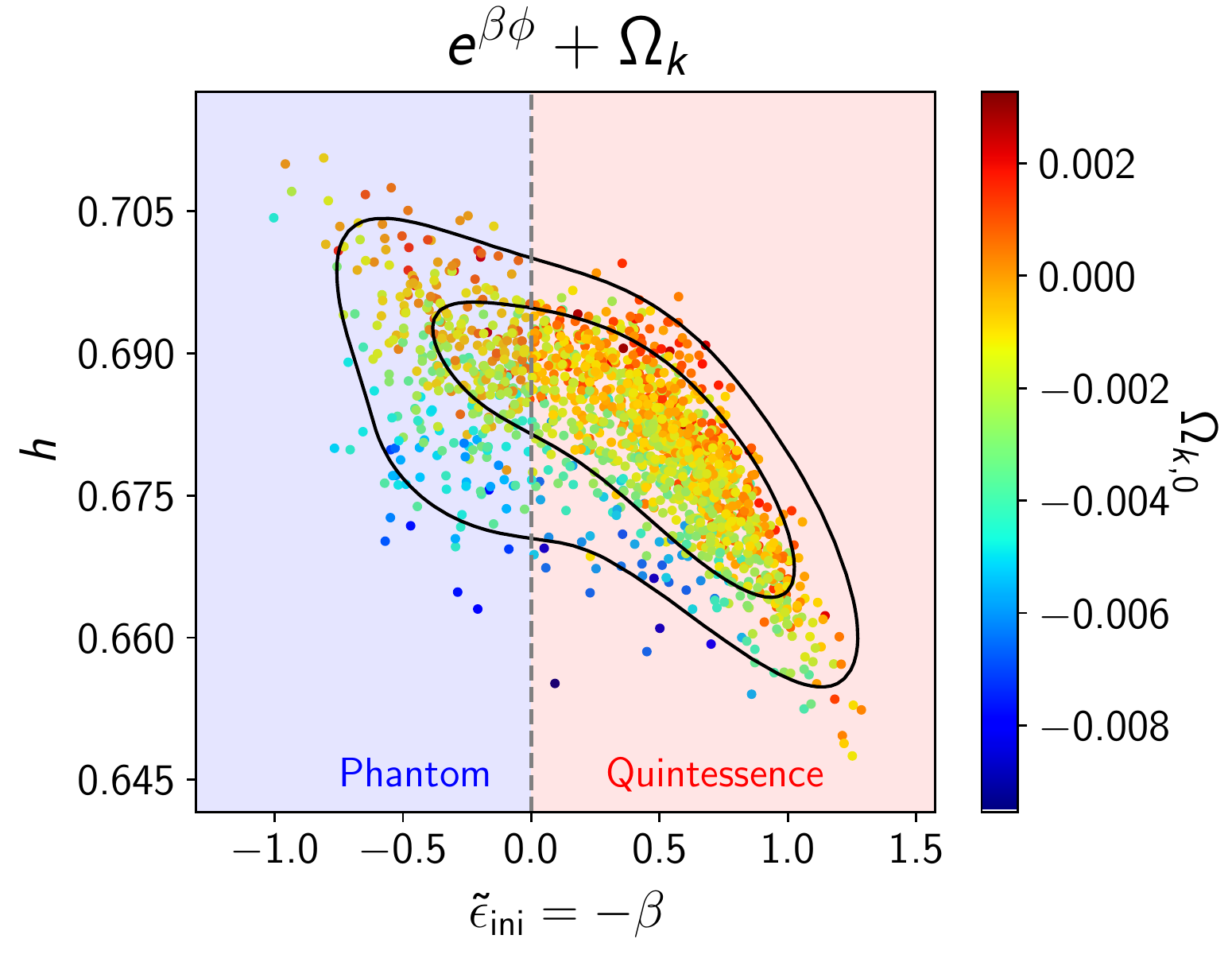} 
\includegraphics[trim = 0mm  0mm 0mm 0mm, clip, width=5.5cm, height=4.5cm]{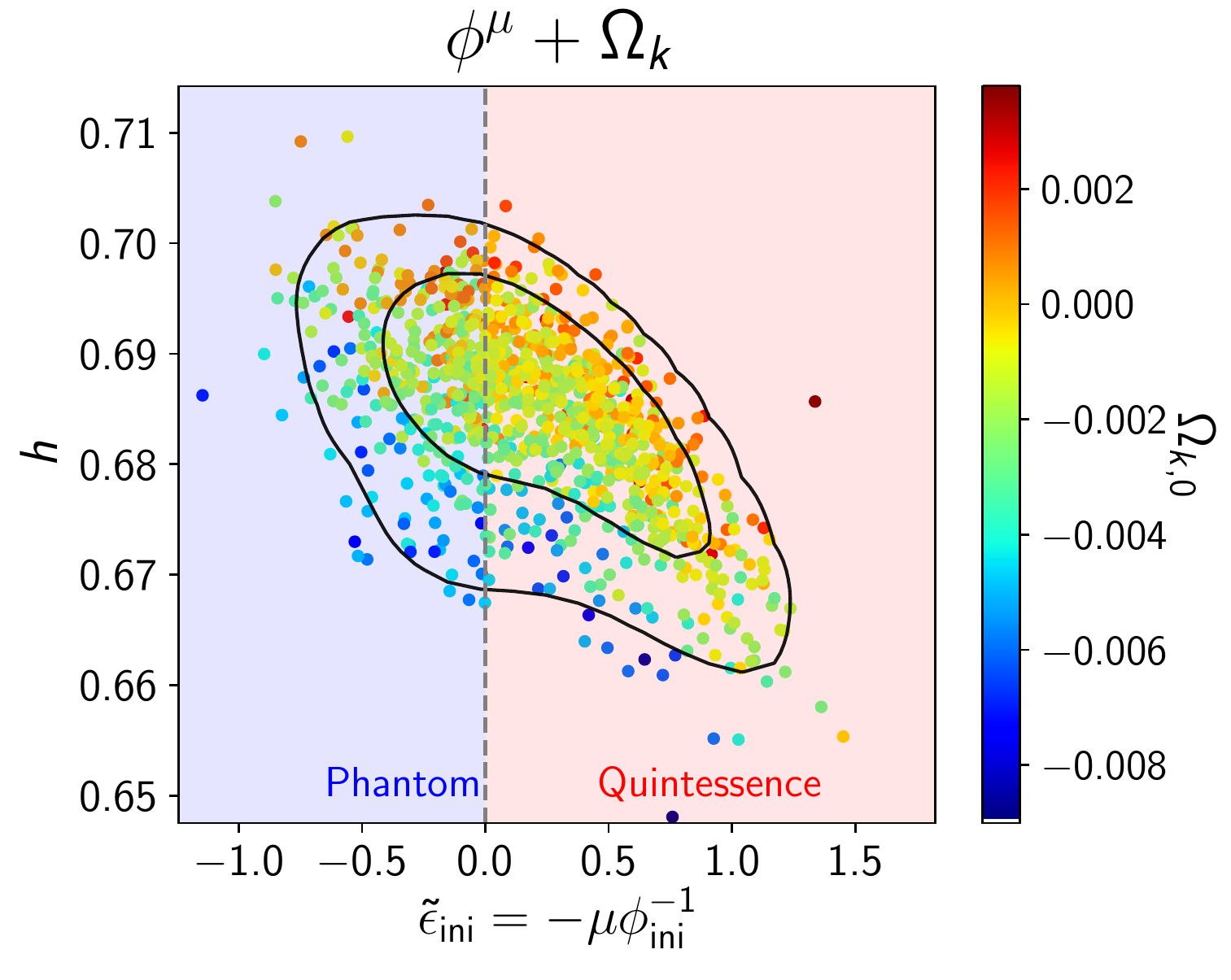} 
\includegraphics[trim = 0mm  0mm 0mm 0mm, clip, width=5.5cm, height=4.5cm]{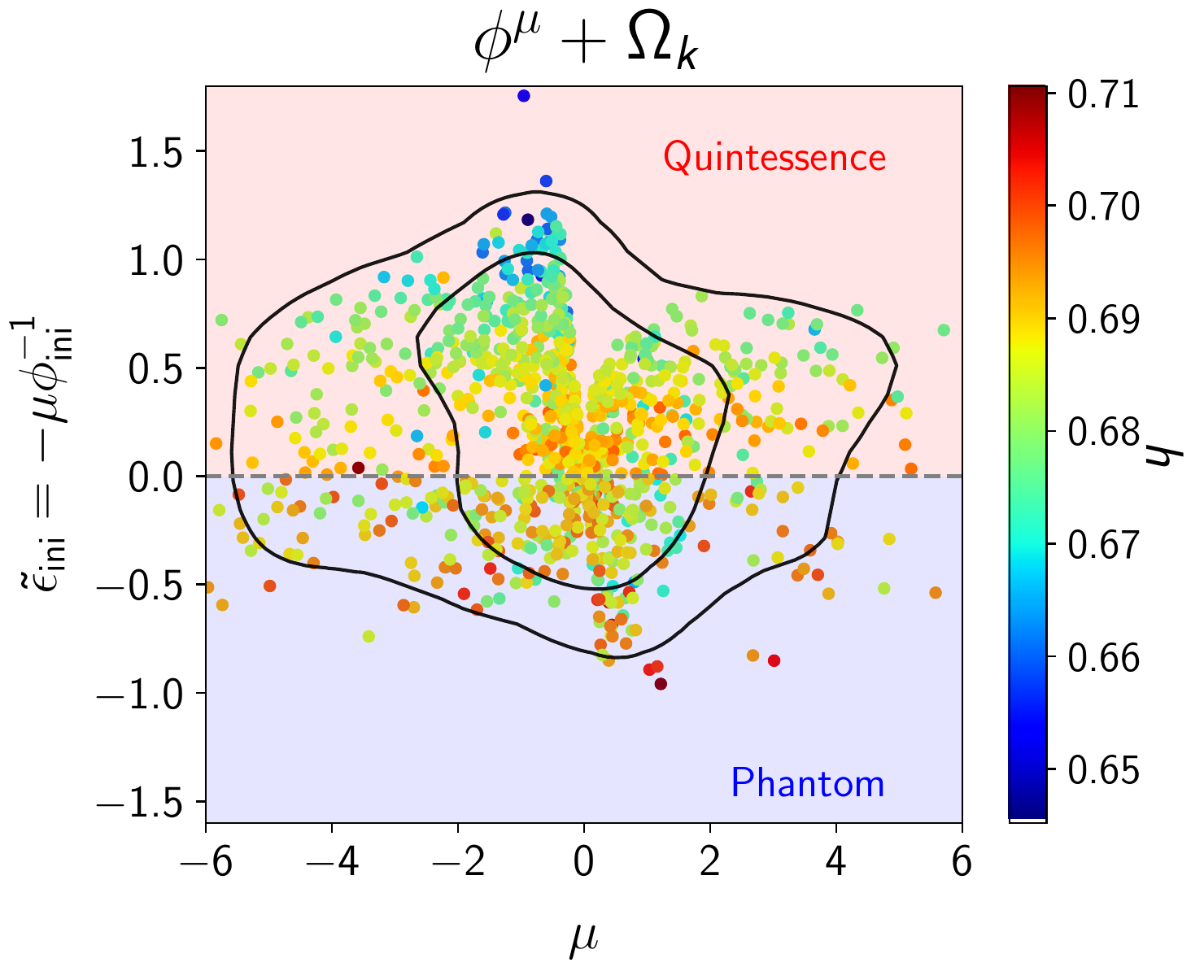} 
\includegraphics[trim = 0mm  0mm 0mm 0mm, clip, width=5.5cm, height=4.5cm]{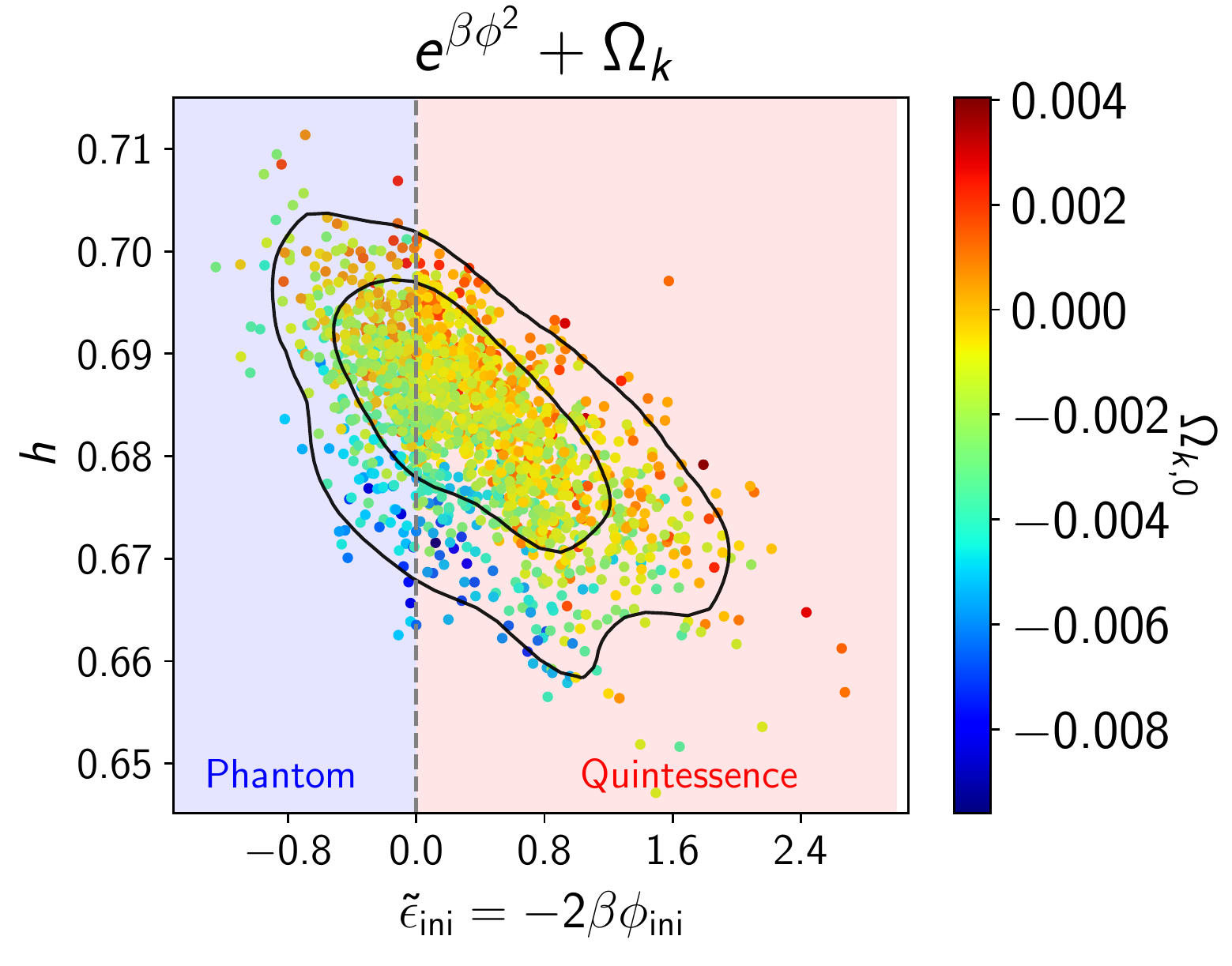} 
\includegraphics[trim = 0mm  0mm 0mm 0mm, clip, width=5.5cm, height=4.5cm]{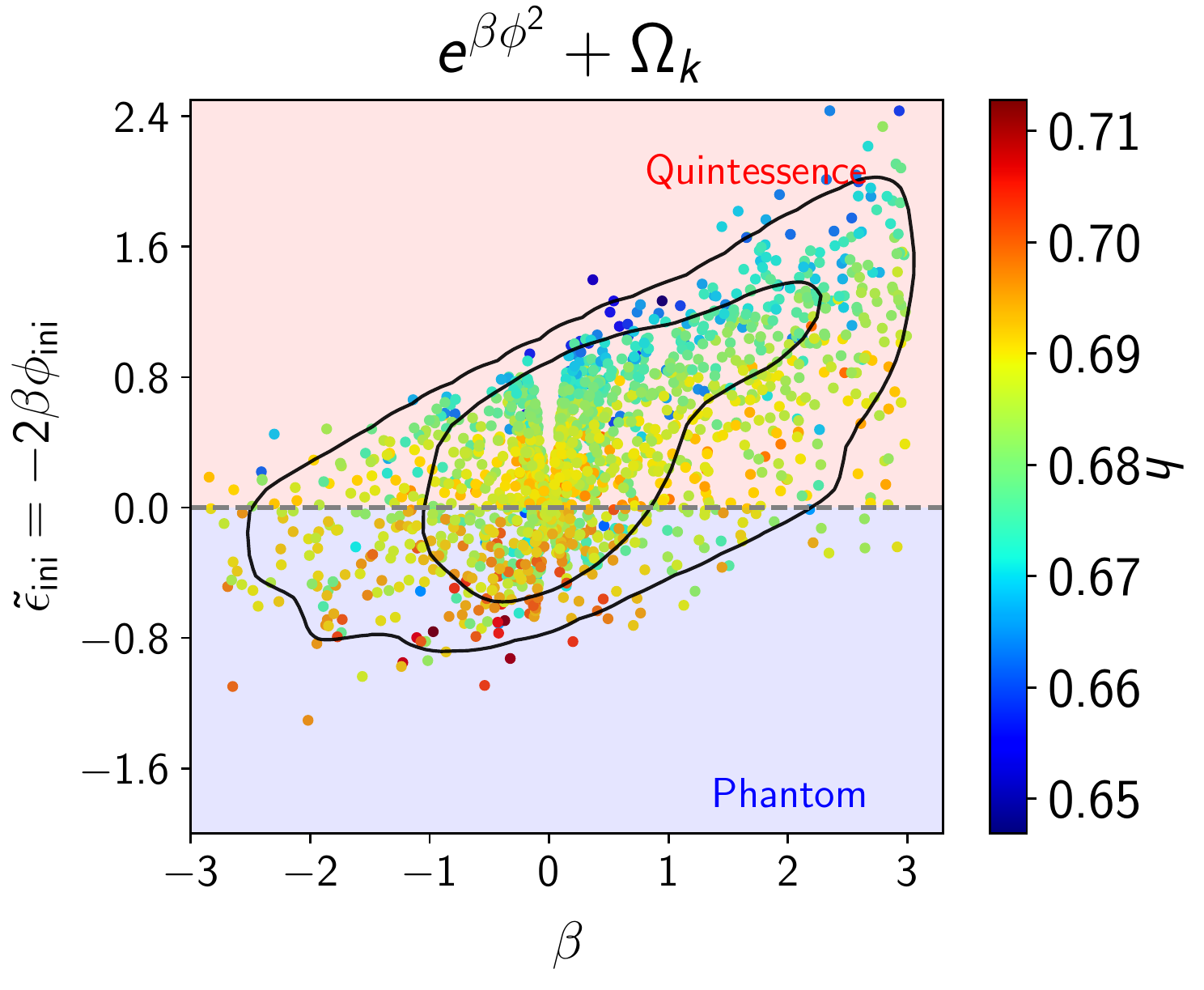} 
\includegraphics[trim = 0mm  0mm 0mm 0mm, clip, width=5.5cm, height=4.5cm]{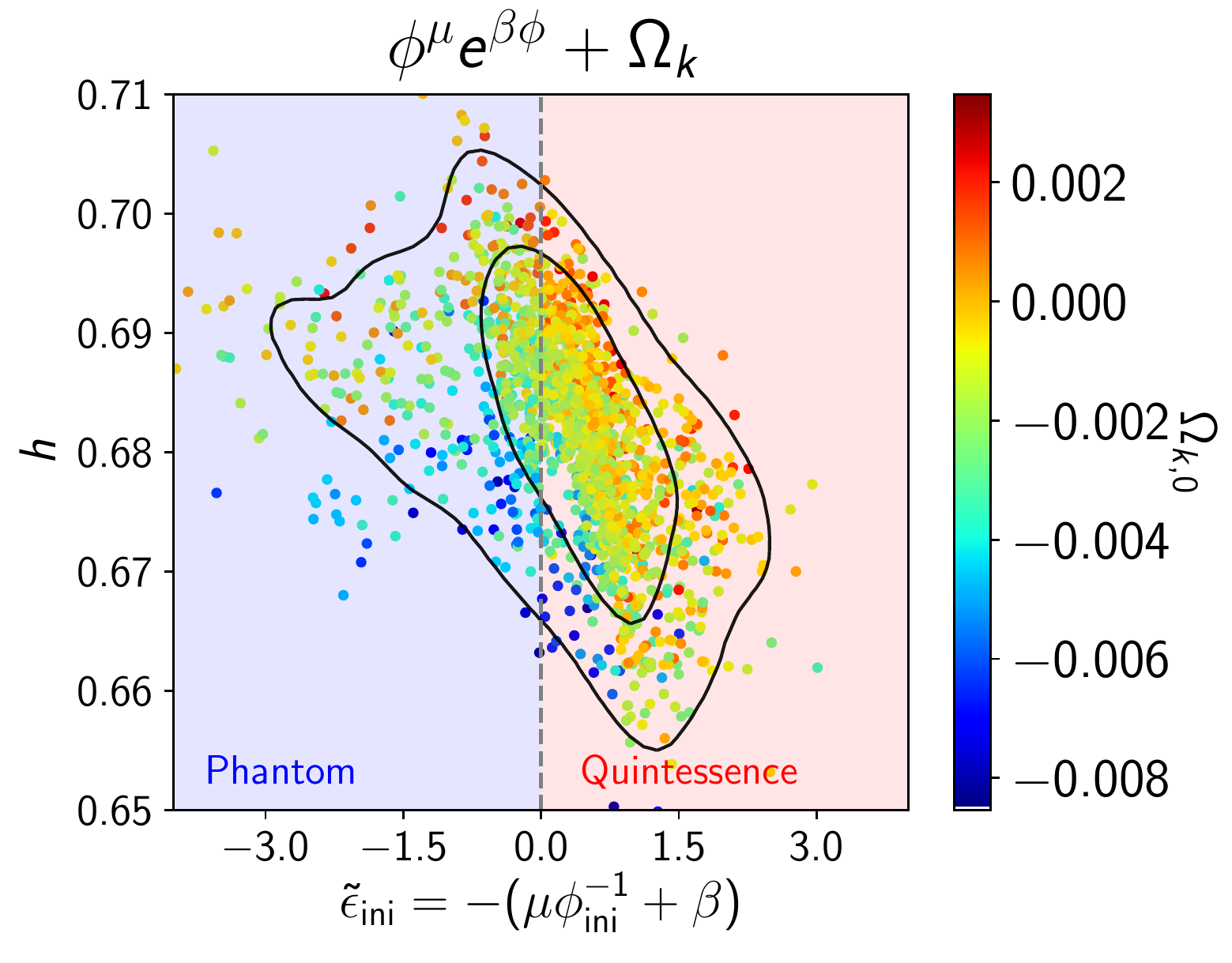}
\includegraphics[trim = 0mm  0mm 0mm 0mm, clip, width=5.5cm, height=4.5cm]{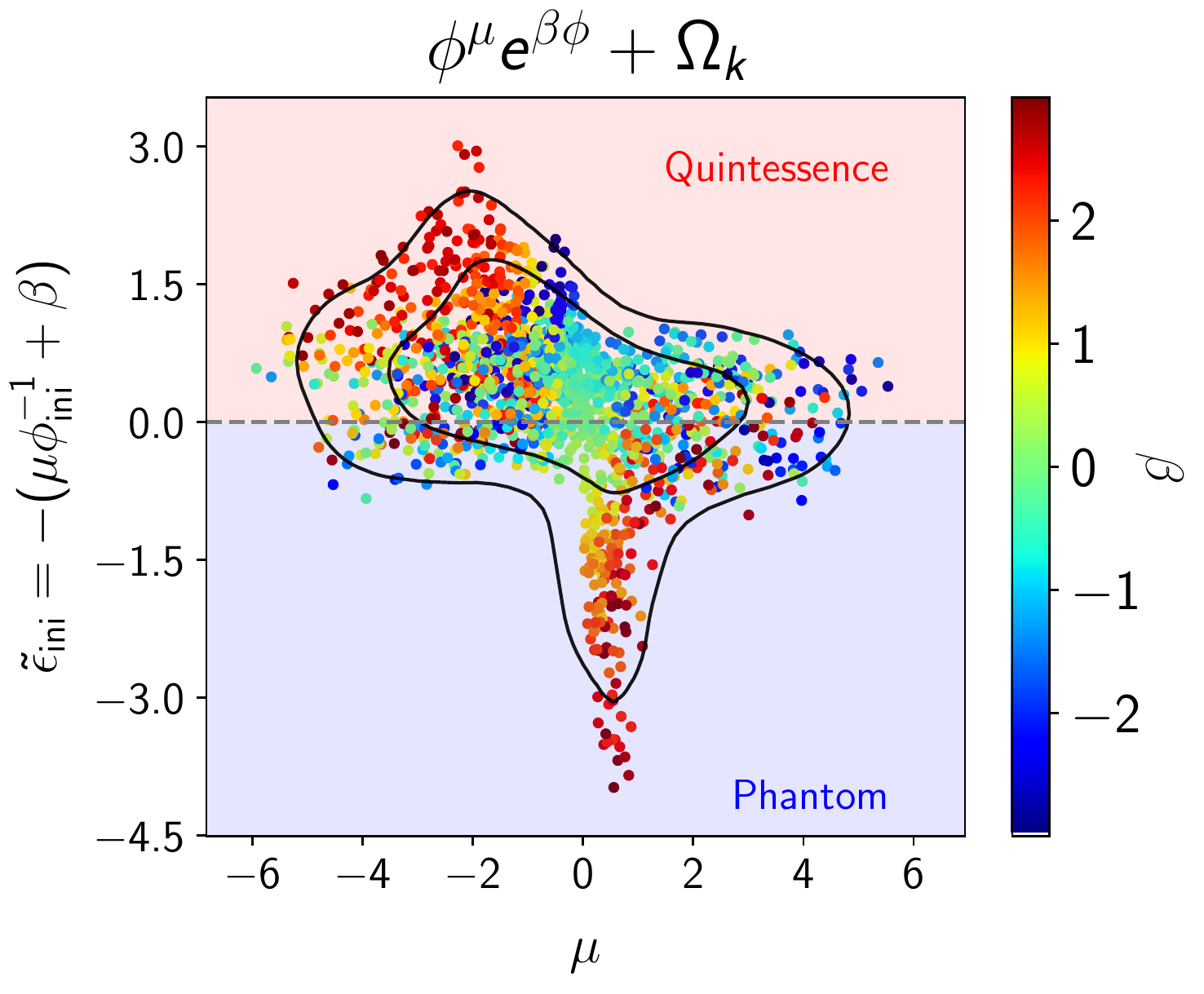} 
\includegraphics[trim = 0mm  0mm 0mm 0mm, clip, width=5.5cm, height=4.5cm]{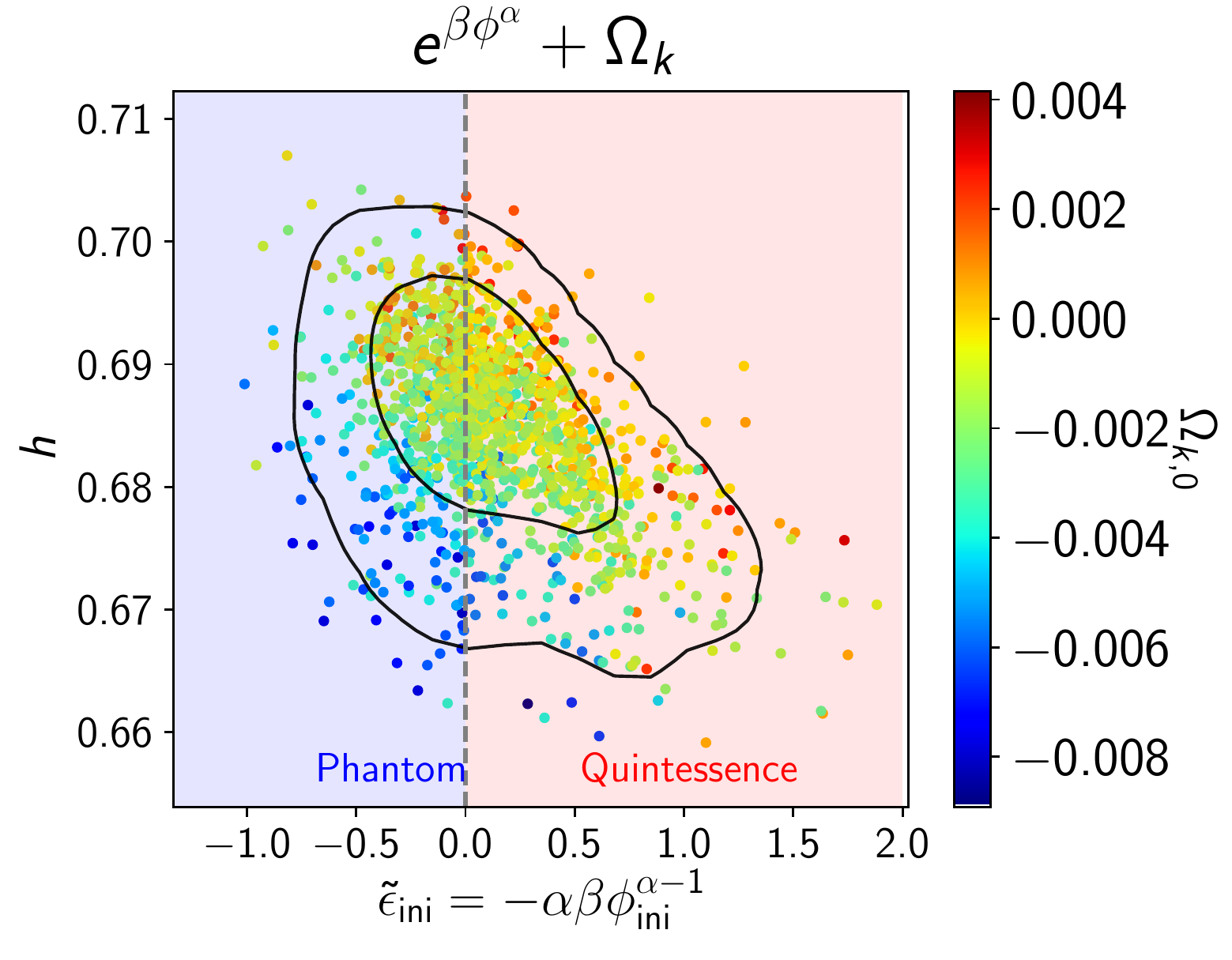} 
\includegraphics[trim = 0mm  0mm 0mm 0mm, clip, width=5.5cm, height=4.5cm]{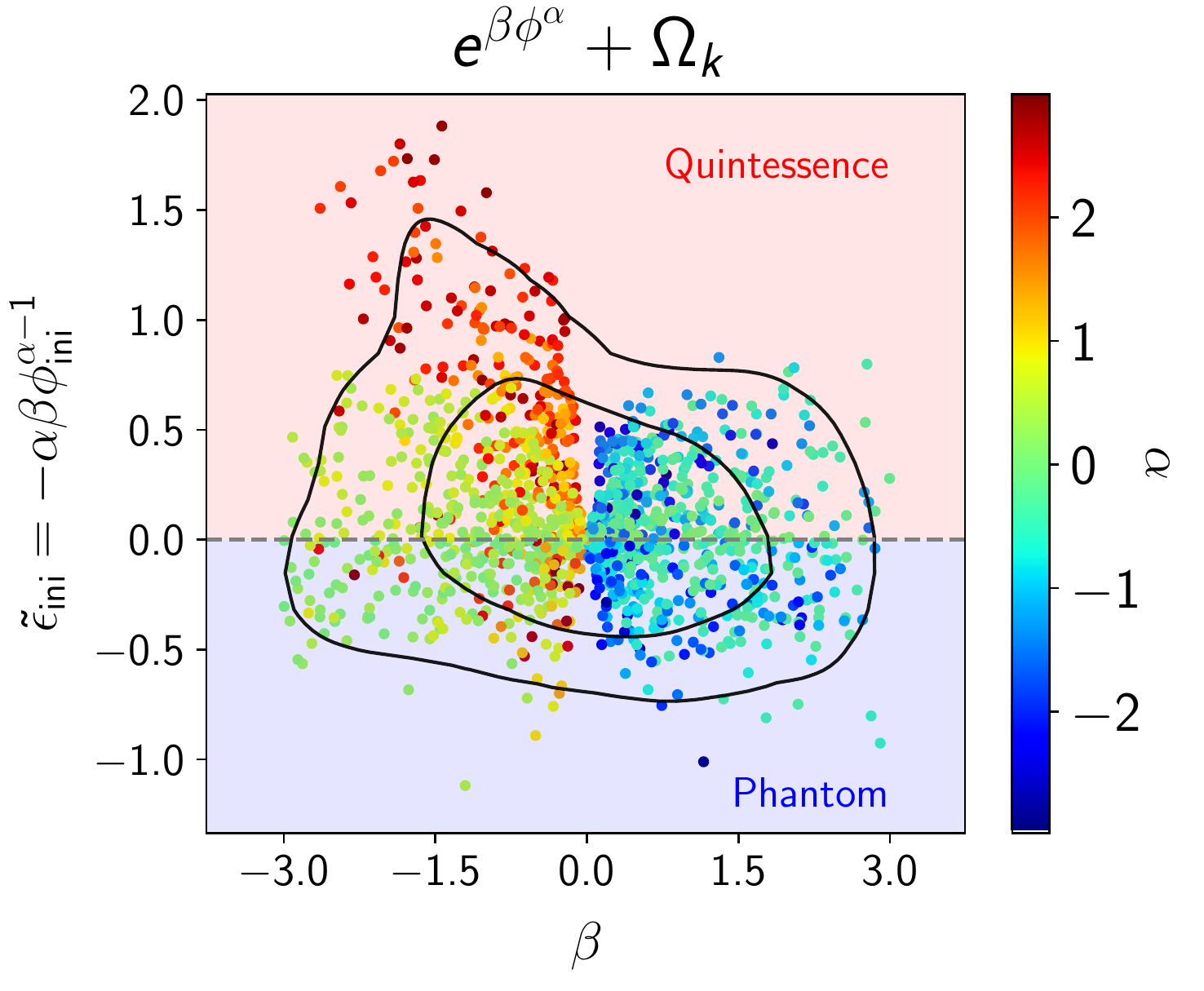} 

\includegraphics[trim = 0mm  0mm 0mm 0mm, clip, width=5.5cm, height=4.5cm]{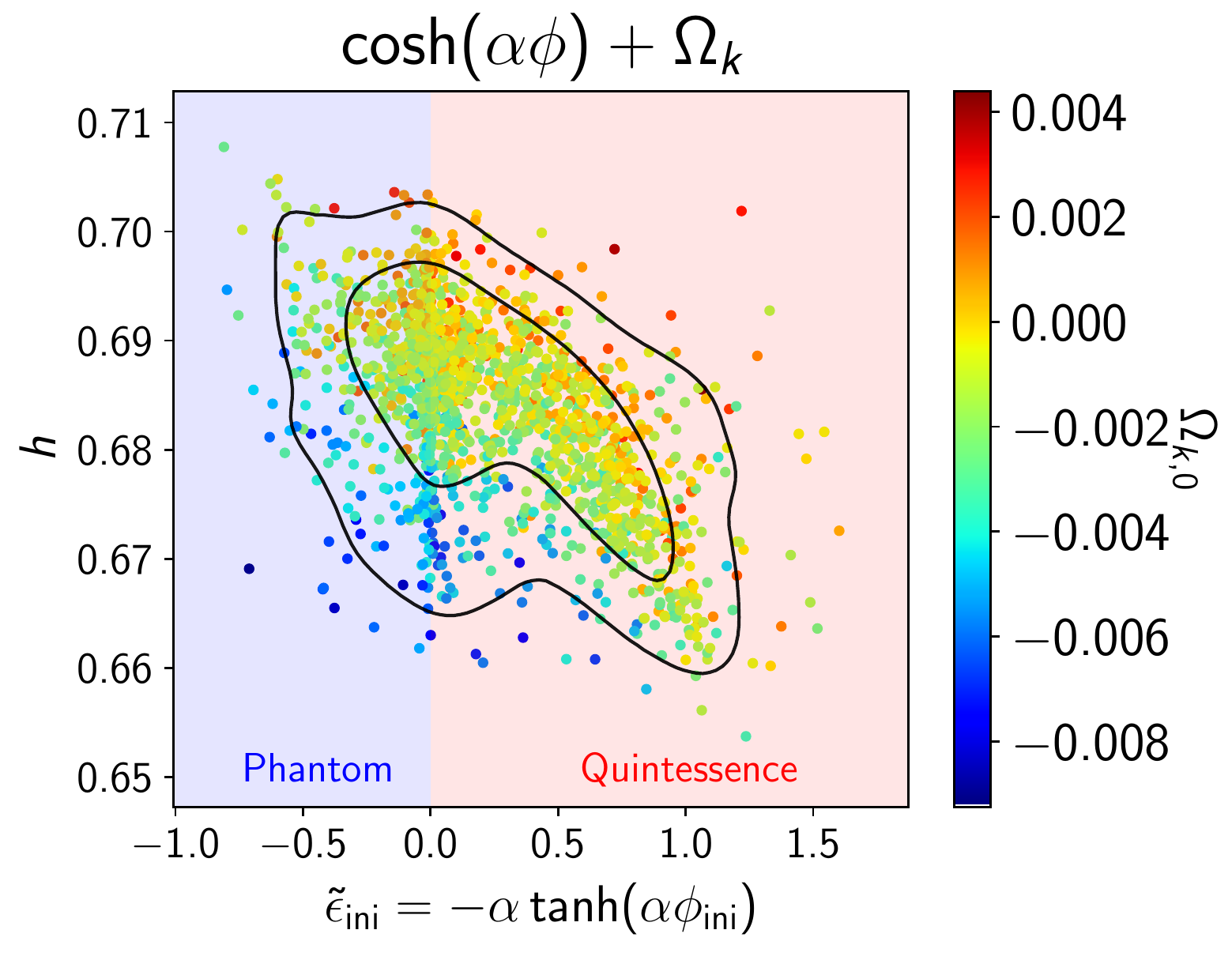} 
\includegraphics[trim = 0mm  0mm 0mm 0mm, clip, width=5.5cm, height=4.5cm]{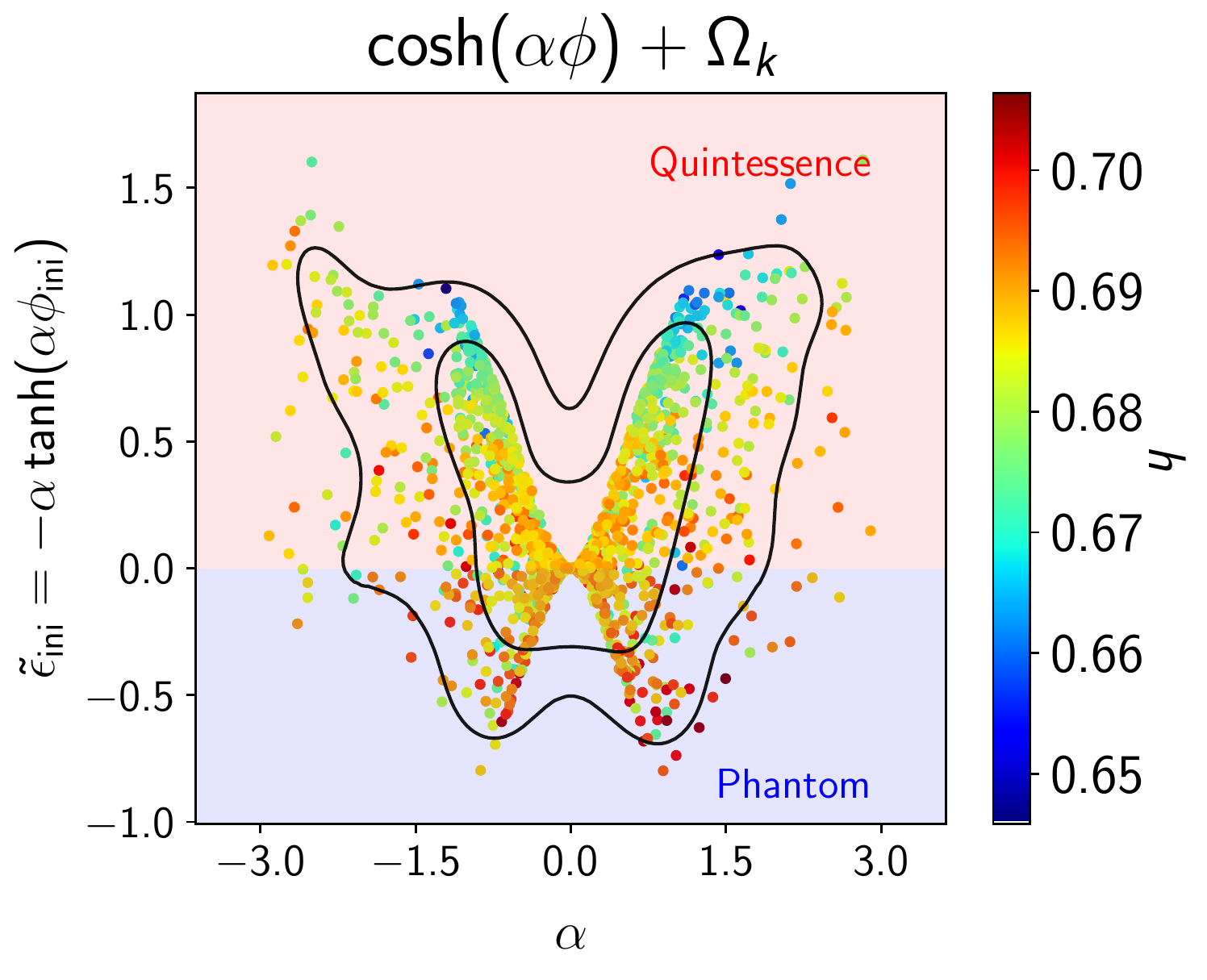} 
\includegraphics[trim = 0mm  0mm 0mm 0mm, clip, width=5.5cm, height=4.5cm]{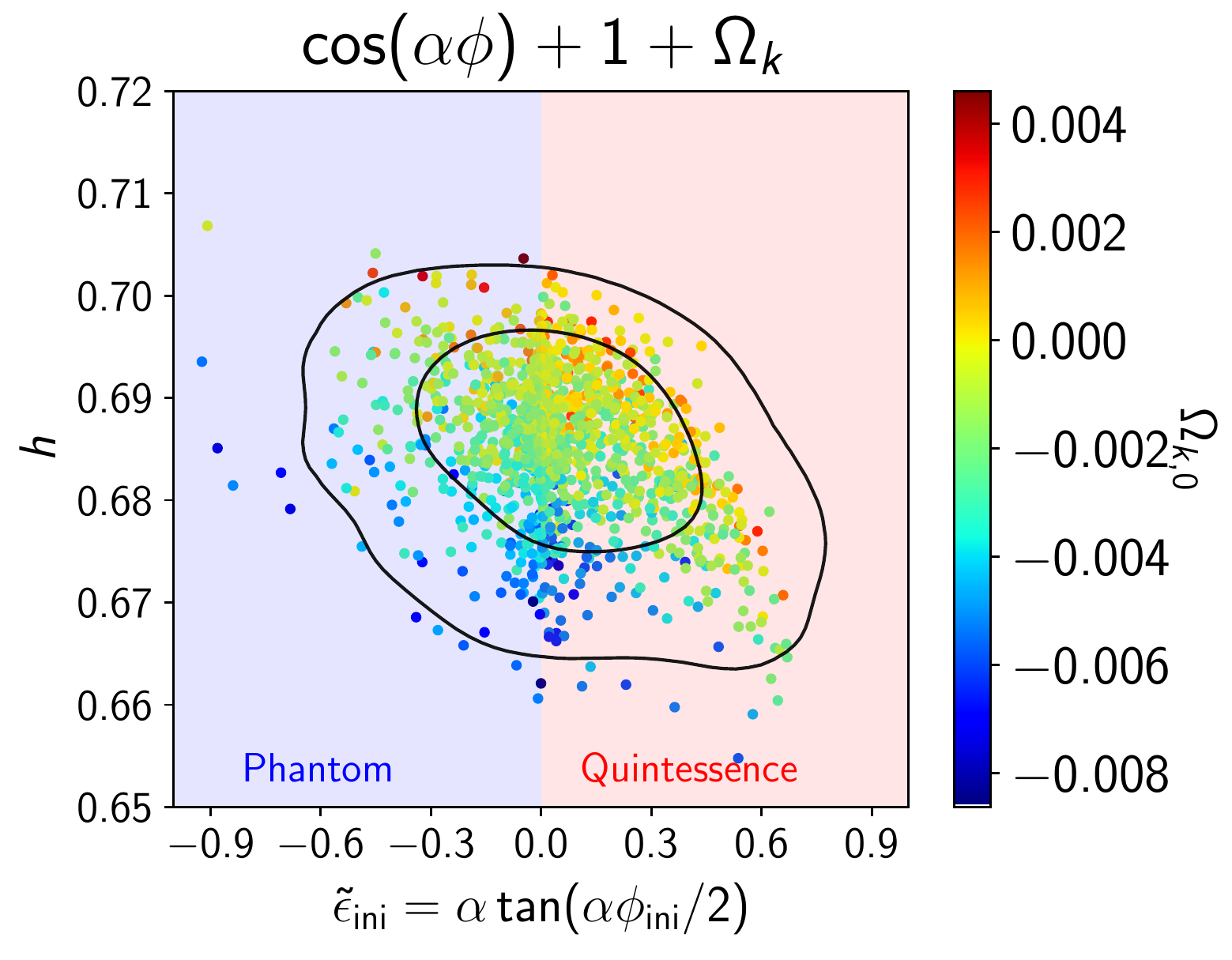}

\includegraphics[trim = 0mm  0mm 0mm 0mm, clip, width=5.5cm, height=4.5cm]{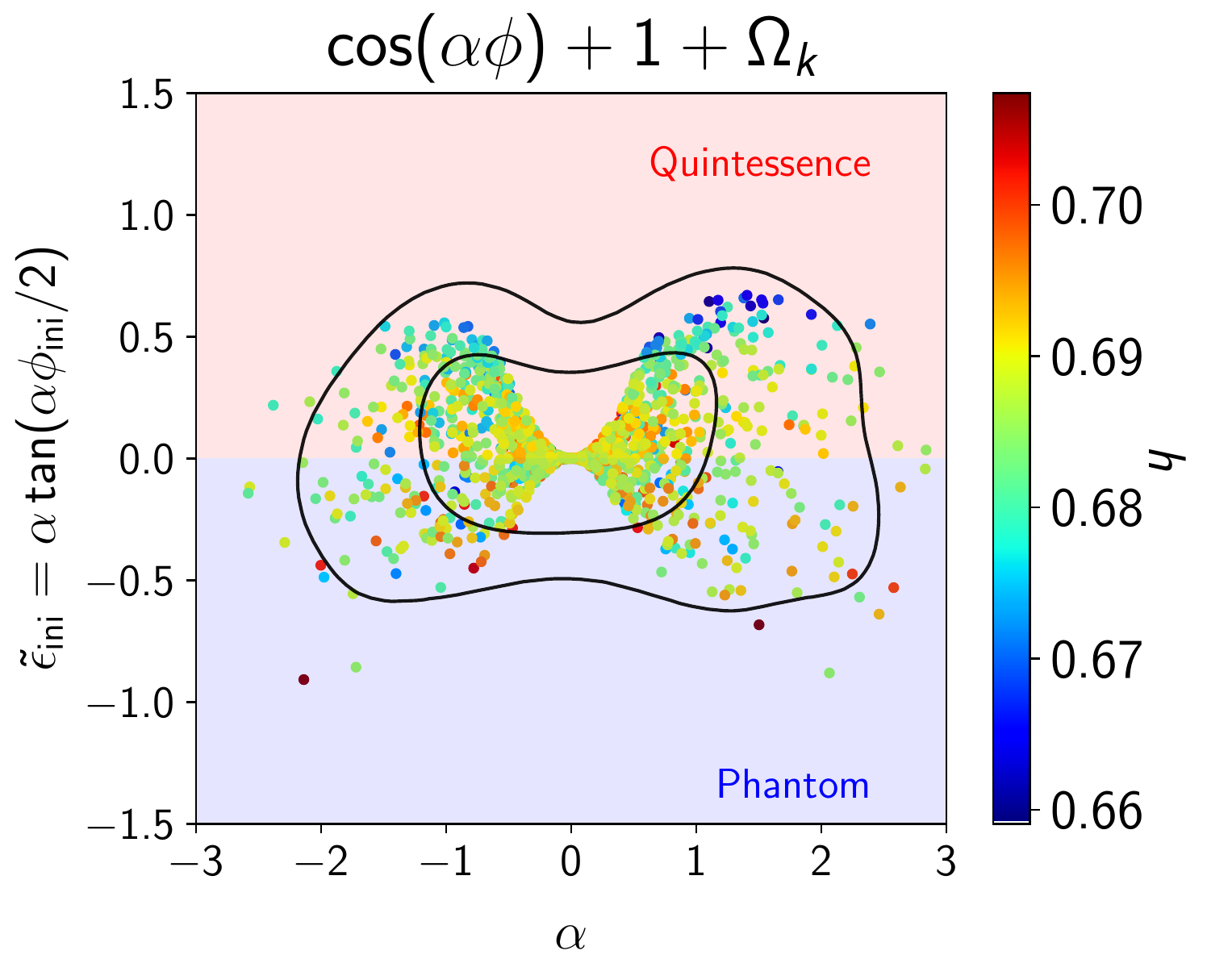} 
\end{center}
\caption[Quintom2]{2D marginalized posterior distributions for each set of parameters give and potential. Colour blue represent models laying on the phantom region, whereas red those models in the quintessence. Inner (outer) contours describe 1$\sigma$ (68\%) and 2$\sigma$ (95\%) confidence contour levels.}
\label{Fig:posteriors}
\end{figure*}

Figure \ref{Fig:posteriors} shows the 2D marginalised posterior distributions with scatter points 
coloured coded accordingly with the parameter on the right coloured bar.  
Notice that each panel contains two colours that represent the type of field: Phantom (blue) and
Quintessence (red), depending on the sign of $\teps_{\rm ini}$ and therefore on the combination 
of parameters for each potential. 
In almost all the presented models, except for the $\cos(\alpha \phi) +1$ model, we notice the centre of the 2D posterior distribution is slightly inclined to the quintessence field.

\begin{figure}[!th]
\includegraphics[trim = 0mm  0mm 0mm 1mm, clip, width=9.cm, height=5.cm]{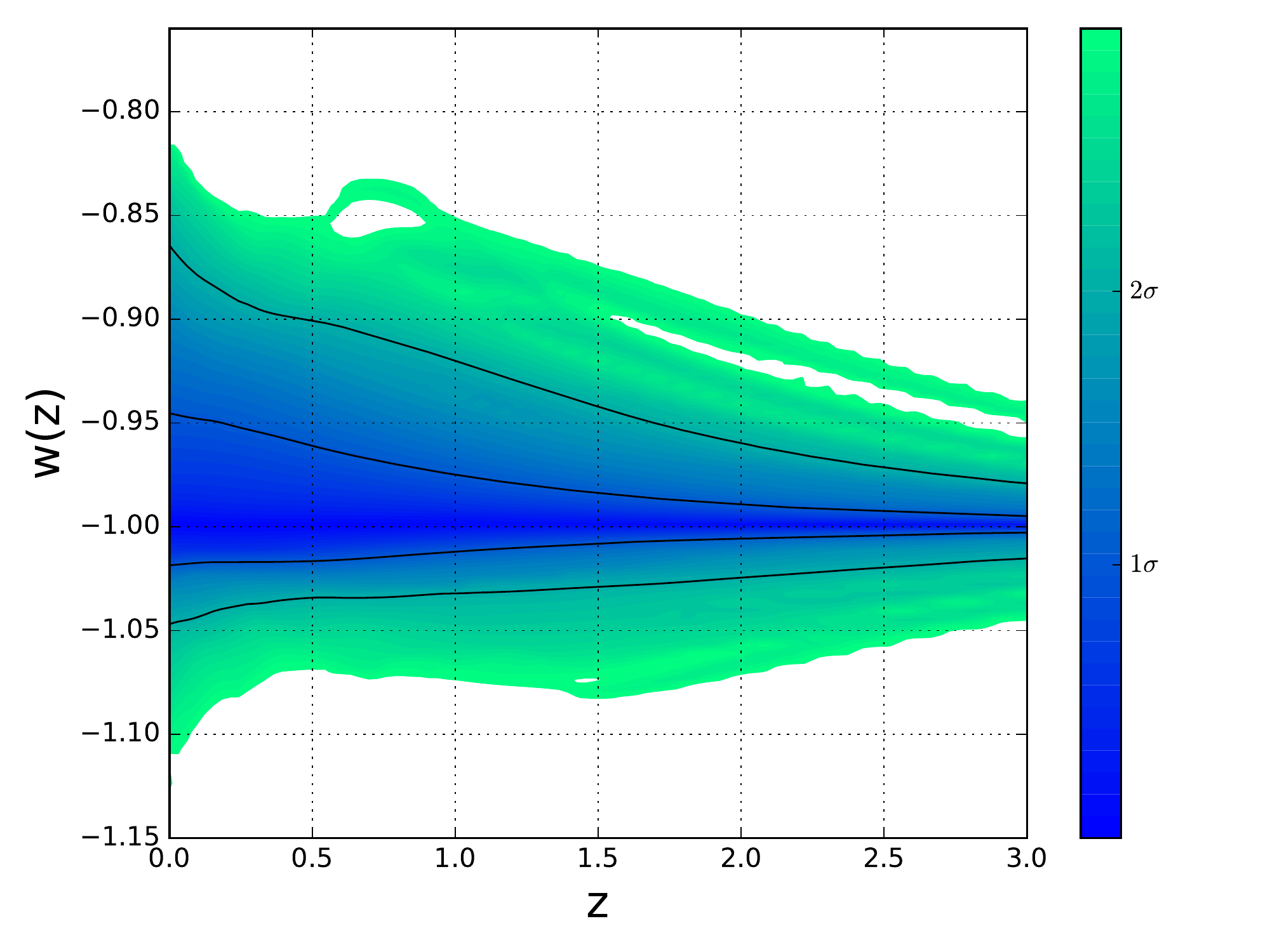} 
\caption{This panel shows the posterior probability $Pr(w|z)$
for the potential that best fits to the data ($V=V_0\phi^\mu e^{\beta \phi}$): the probability of $w$ 
as normalised in each slice of constant $z$, with colour scale in confidence 
interval values.  
Solid lines describe 1$\sigma$ (68\%) and 2$\sigma$ (95\%) confidence contour levels.}
\label{Fig:posterior_w}
\end{figure}

Within the potentials studied, the mean value of the Hubble parameter is around $H_{0} = 68 km/s/Mpc$ which is consistent with the value of the $\Lambda$CDM model.
Despite the diversity of potentials and even if the phantom regime is considered, there is not a significant departure from the standard background cosmology that may allow to alleviate the $H_{0}$ tension. If we see Figure \ref{Fig:posterior_w}, up-to $2\sigma$, the $w(z=0)$ can go at most till $-1.1$ which is insufficient to shift the central value of $H_{0}$, as has been shown in \cite{Vagnozzi:2019ezj} that it is needed $w(z=0)$ around $-1.3$ to shift the central value of $H_{0}$ towards $70-71$.

Finally, an important point to stress out is the presence of pronounced degeneracies over the parameter-space. 
For instance, in the model with potential $V(\phi)=V_0 e^{\beta \phi^\alpha}$ the parameter space is divided
into two regions, one with $\beta>0, \alpha<0$ and the other one with $\beta<0, \alpha>0$ and in the
centre of this region $V(\phi)=V_0$. Something similar happens with the potential 
$V(\phi)=V_0\phi^\mu e^{\beta \phi}$. 
A notable feature for this case is the fact that the points do not cluster near the centre of the confidence curve, and also clearly the particular case $\beta = 0$ is constrained by the initial condition to $\teps_{ini} \sim 0$.
These type of degeneracies are very complex and time consuming to explore within the standard MCMC methods, hence the use of nested samplers.
To complement this figure, Table \ref{TableII} contains mean values along with $1\sigma$ 
constraints on the set of parameters used to described each model.  
Looking at the mean values of $\teps$, we obtain that in general there is a slight inclination to quintessence models ($\teps >0$) for a open universe ($\Omega_{k,0}<0$) (also observed in Figure \ref{Fig:posteriors}).

In terms of particular models;
the potential $V=V_0e^{\beta \phi}$ slightly favours $\beta$ negative (positive for quintessence and negative for phantom values);
in contrast, the potential $V=V_0e^{\beta \phi^2}$ slightly favours $\beta$, which means there is a maximum in the equation of state.
For $V=V_0e^{\beta \phi^\alpha}$ the parameters $\beta \approx 0.16$ and $\alpha \approx 0.1$; all either for flat or curved Universe; notice that the power $\alpha \approx 0.1$ of the last case lies between the powers 0 and 2 of the first and second cases.
For the power-law potential, $V=V_0\phi^\mu$, models with an exponent $|\mu|>2$ lay down right on the $1\sigma$ limits. 
The model with a potential $V=V_0\phi^\mu e^{\beta \phi}$, the means are located at $\beta \sim 0.2$ and $\mu \sim -0.4$, this potential with negative curvature provides a better fit to the data compared to the rest of the models and shows and improvement of $1.6\sigma$ with respect to the $\Lambda$CDM model, however with three extra parameters.
The potential $V=V_0 \cosh(\alpha \phi)$ has the feature that the best-fit of the parameter $\alpha$ is negative in the flat case and positive in a universe with curvature case with a slightly preference to an open universe.
Finally for the potential $V=V_0(\cos(\alpha \phi) + 1)$, the parameter $\alpha \sim 0$ and $\alpha \sim 0.4$ for the flat and curvature cases respectively.

Last column contains the Bayesian evidence ($\ln \mathcal{Z}$), and according to Jeffrey's scale, the models are indistinguishable and hence still consistent with $\Lambda$CDM.
In general we found that scalar field models preferred an open universe compared to positive curvature for the $\Lambda$CDM model.

\begin{table*}[ht!]
\footnotesize
\scalebox{0.99}{%
\begin{tabular}{ccccccccccc} 
\cline{1-11}\noalign{\smallskip}
 \vspace{0.15cm}

  Model & $\Omega_{\rm m, 0}$ & $h$  & $\Omega_{k,0}$  & $\phi^{-1}_{\rm ini}$  & $\beta$ & $\mu$ & $\alpha$&  $\teps_{\rm ini}$&
  $-2\ln \mathcal{L}_{\rm max}$ &  $\ln \mathcal{Z}$  \\
\hline
\vspace{0.15cm}
$\Lambda$CDM   &$0.3005(68)$ &$0.6830(53)$ &0&$-$  & $-$& $-$ &$-$ & $-$& $60.44$ &  $-40.39(20)$\\
\vspace{0.15cm}
$\Lambda$CDM $+k$          &$0.3008(67)$ & $0.6849(65)$ &$0.0013(18)$ & $-$ 
 & $-$& $-$ & $-$&  $-$&  $59.30$ &  $-41.10(22)$\\
\vspace{0.15cm}
 $e^{\beta \phi}$   &$0.3023(91)$ &$0.6838(98)$ &0& $-$  & $-0.37(43)$ & $0$ & $1$
  & $0.37(43)$ & $59.00$ &  $-40.81(21)$\\
\vspace{0.15cm}
$e^{\beta \phi} + k$     &$0.3026(86)$ & $0.6812(95)$ & $-0.0016(17)$  & $-$
  &$-0.34(44)$ & $0$ & $1$ & $0.34(44)$ & $58.46$ &  $-42.02(23)$\\
\vspace{0.15cm}
 $\phi^\mu$           &$0.3009(79)$ & $0.6856(85)$ & $0$   & $0.15(91)$ & $0$
& $-0.13(2.13)$ & $-$ & $0.29(42)$ &  $58.94$& $-40.95(22)$\\
\vspace{0.15cm}
 $\phi^\mu + k$       &$0.3001(78)$ &$0.6842(83)$ & $-0.0016(18)$   & $0.31(96)$
 & $0$ & $-0.31(1.88)$ & $-$ & $0.22(41)$ &  $58.32$ & $-42.05(23)$ \\
\vspace{0.15cm}
 $e^{\beta \phi^2}$     &$0.3003(78)$  & $0.6863(83)$  &$0$& $-0.13(96)$  &$0.39(1.07)$ 
 & $0$ &$2$   & $0.36(55)$& $58.32$ &  $-41.36(22)$\\ 
\vspace{0.15cm}
 $e^{\beta \phi^2}+k$   &$0.3011(80)$  & $ 0.6832(87)$  & $-0.0015(17)$  &$-0.17(99)$ 
 & $0.34(1.11)$& $0$  &$2$& $0.34(56)$& $58.18$ &  $-42.18(23)$\\ 
\vspace{0.15cm}
 $\phi^\mu e^{\beta \phi}$  &$0.3013(82)$  & $0.6850(90)$   &$0$ & $0.23(1.29)$ &  
 $0.25(1.52)$ & $-0.37(2.17)$& $1$  &$0.32(90)$&  $58.44$ & $-41.41(22)$ \\ 
\vspace{0.15cm}
 $\phi^\mu e^{\beta \phi}+k$  &$0.3019(82)$  & $0.6820(90)$   &$-0.0016(18)$ & $0.27(1.31)$ 
 & $0.22(1.56)$ &  $-0.41(1.90)$ &  $1$ & $0.26(92)$& $58.02$ & $-42.42(23)$ \\ 
\vspace{0.15cm}
 $e^{\beta \phi^\alpha}$    &$0.2989(73)$  & $0.6879(75)$   &$0$ & $0.18(0.93)$ & $-0.04(1.27)$ 
 & $0$  & $0.11(1.28)$& $0.19(42)$& $58.48$ & $-41.67(22)$\\ 
\vspace{0.15cm}
 $e^{\beta \phi^\alpha}+k$    &$0.2993(71)$  & $0.6853(74)$   &$-0.0017(18)$ & $0.15(1.01)$ & 
 $-0.05(1.15)$ & $0$  & $0.09(1.22)$& $0.12(39)$& $58.22$ & $-42.57(23)$\\ 
\hline
\vspace{0.15cm}
 $\cosh(\alpha \phi)$    &$0.30087(80)$  & $0.6858(85)$ &$0$  &$-0.31(1.00)$ & $0$  
 & $-$  & $-0.61(1.36)$& $0.34(46)$& $58.50$ & $-40.63(21)$\\ 
 \vspace{0.15cm}
 $\cosh(\alpha \phi) + k$    &$0.30090(79)$  & $0.6835(85)$ &$-0.0016(18)$  &$-0.43(1.31)$ 
 & $0$  & $-$  & $0.02(93)$& $0.26(41)$& $58.14$ & $-41.47(22)$\\ 
 \vspace{0.15cm}
 $\cos(\alpha \phi) +1$   &$0.2977(66)$  & $0.6895(61)$ &$0$  &$0.23(1.41)$ 
 & $1$  & $-$  & $0.01(1.03)$& $0.06(69)$& $59.12$ & $-40.76(21)$\\ 
   \vspace{0.15cm}
 $\cos(\alpha \phi) +1 + k$    &$0.2992(69)$  & $0.6855(73)$ &$-0.0018(18)$  &$0.07(0.94)$ 
 & $1$  & $-$  & $0.38(1.33)$& $0.14(58)$& $58.54$ & $-41.48(22)$\\ 
\hline
\end{tabular}}
\caption{ Mean values along with $1\,\sigma$ constraints on the set of parameters used to described each model. 
For one-tailed distributions the upper limit 
95\% C.L. is given. For two-tailed the 68\% C.L. is shown. 
Before the last column, $-2\ln \mathcal{L}_{{\rm,max}}$ 
is used to compare best-fit with respect to the $\Lambda$CDM
model. Last column contains the Bayesian evidence.}
\label{TableII}
\end{table*}

\section{Conclusions}\label{Conclusions}

In this paper we have been able to incorporate minimally coupled scalar fields 
- quintessence and phantom -  
into the same analysis with the use of a switch parameter $\teps$, 
and then to present it as a python module for the SimpleMC code \cite{SimpleMC}. 
The variables introduced here allow us to write down the dynamical system in terms 
of variables that naturally simplify the selection of initial conditions.  
This $\teps$ parameter will be useful when the two fields are combined, i.e. quintom models.
For the proof of the concept we have considered two generic classes of potentials, a three-parametric class $V(\phi)= V_0 \phi^\mu e^{\beta \phi^\alpha}$ and also the two-parametric class $V(\phi)= V_0(\cosh(\alpha \phi) + \beta$), 
but the code can easily be extended to more specific potentials. 
One of the objectives of the code is to enable the study of very general models of scalar dark energy potentials such as the one presented in this work.
We found that for the combined dataset, the preferred model corresponds to 
$V=V_0\phi^\mu e^{\beta \phi}$ and negative curvature.
For this potential, Figure \ref{Fig:posterior_w} contains the posterior probability of the equation of state $Pr(w|z)$: 
the probability of $w$ as normalised in each slice of constant $z$, with colour scale in confidence interval values \cite{Handley}. 
The Bayesian evidence points out an inconclusive difference
amongst models. For the coming datasets this work may be able to identify the type of field and the scalar field potential that best describes observations.

\begin{acknowledgments}
JAV acknowledges the support provided by FOSEC SEP-CONACYT Investigaci\'on B\'asica A1-S-21925, FORDECYT-PRONACES/304001/2020 and UNAM-DGAPA-PAPIIT  IA102219.
DT acknowledges the receipt of the grant from the Abdus Salam International Centre for Theoretical Physics, Trieste, Italy; and would like to thank Karen Caballero for the hospitality provided at the Facultad de Ciencias en F\'isica y Matem\'aticas, Universidad Aut\'onoma de Chiapas.
AAS acknowledges funding from DST-SERB, Govt of India, under the project NO. MTR/20l9/000599.

\end{acknowledgments}


\begin{thebibliography}{99}

\bibitem{Huterer:2017buf}
  D.~Huterer and D.~L.~Shafer,
  Rept.\ Prog.\ Phys.\  {\bf 81} (2018) no.1,  016901
  doi:10.1088/1361-6633/aa997e
  [arXiv:1709.01091].

\bibitem{lambda-p1} P. J. E. Peebles, B. Ratra, 
Astrophys. J. Lett. {\bf 325} (1988) L17

\bibitem{lambda-p2} S. Weinberg, 
,Rev. Mod. Phys. {\bf 61} (1989) 1-23
  
\bibitem{lambda-p3} A. G. Riess et al (Supernova Search Team), 
Astron. J. {\bf 116} (1998) 1009-1038 [e-Print:astro-ph/9805201].


\bibitem{Zlatev:1998tr}
I.~Zlatev, L.~M.~Wang and P.~J.~Steinhardt,
Phys. Rev. Lett. \textbf{82} (1999), 896-899
doi:10.1103/PhysRevLett.82.896
[arXiv:astro-ph/9807002].

\bibitem{lambda-p5} S. M. Carroll, 
Living Rev. Rel. {\bf 4} (2001) 1 [e-Print:astro-ph/0004075].

\bibitem{lambda-p6} P. J. E. Peebles, B. Ratra, 
Rev. Mod. Phys. {\bf 75} (2003) 559-606 [e-Print:astro-ph/0207347].

\bibitem{lambda-p7} T. Padmanabhan, 
Phys. Rept. {\bf 380} (2003) 235-320 [e-Print:hep-th/0212290].

\bibitem{Vazquez:2012ce}
  J.~Alberto Vazquez, M.~Bridges, M.~P.~Hobson and A.~N.~Lasenby,
  JCAP {\bf 1209} (2012) 020
  doi:10.1088/1475-7516/2012/09/020
  [arXiv:1205.0847].
  
\bibitem{Hee:2016nho}
  S.~Hee, J.~A.~V\'azquez, W.~J.~Handley, M.~P.~Hobson and A.~N.~Lasenby,
  Mon.\ Not.\ Roy.\ Astron.\ Soc.\  {\bf 466} (2017) no.1,  369
  doi:10.1093/mnras/stw3102
  [arXiv:1607.00270].
  
\bibitem{Zhao:2017cud}
  G.~B.~Zhao {\it et al.},
  Nat.\ Astron.\  {\bf 1} (2017) no.9,  627
  doi:10.1038/s41550-017-0216-z
  [arXiv:1701.08165].

\bibitem{Wang:2018fng}
  Y.~Wang, L.~Pogosian, G.~B.~Zhao and A.~Zucca,
  Astrophys.\ J.\  {\bf 869} (2018) L8
  doi:10.3847/2041-8213/aaf238
  [arXiv:1807.03772].
  
\bibitem{Tamayo:2019gqj}
Tamayo, David and Vazquez, J. Alberto,
Mon. Not. Roy. Astron. Soc., 487, 1, 729--736, 2019
doi :10.1093/mnras/stz1229,
[arXiv:1901.08679].

\bibitem{Yang:2018qmz}
W.~Yang, S.~Pan, E.~Di Valentino, E.~N.~Saridakis and S.~Chakraborty,
Phys. Rev. D \textbf{99} (2019) no.4, 043543
doi:10.1103/PhysRevD.99.043543
[arXiv:1810.05141].

\bibitem{Pan:2019gop}
S.~Pan, W.~Yang, E.~Di Valentino, E.~N.~Saridakis and S.~Chakraborty,
Phys. Rev. D \textbf{100} (2019) no.10, 103520
doi:10.1103/PhysRevD.100.103520
[arXiv:1907.07540].

\bibitem{Tsujikawa:2013fta}
S.~Tsujikawa,
Class. Quant. Grav. \textbf{30} (2013), 214003
doi:10.1088/0264-9381/30/21/214003
[arXiv:1304.1961].

\bibitem{quint-1} S. M. Carroll, 
Phys. Rev. Lett. {\bf 81} (1998) 3067-3070 [e-Print:astro-ph/9806099].

\bibitem{quint-2} L. M. Wang, P. J. Steinhardt, 
Astrophys. J. {\bf 508} (1998) 483-490 [e-Print:astro-ph/9804015].

\bibitem{quint-3} L. Amendola, 
Phys. Rev. D {\bf 62} (2000) 043511 [e-Print:astro-ph/9908023].

\bibitem{quint-4} T. Chiba, T. Okabe, M. Yamaguchi, 
Phys. Rev. D {\bf 62} (2000) 023511 [e-Print:astro-ph/9912463].

\bibitem{quint-5} L. M. Wang, R. R. Caldwell, J. P. Ostriker, P. J. Steinhardt, 
Astrophys. J. {\bf 530} (2000) 17-35 [e-Print:astro-ph/9901388].

\bibitem{quint-6} W. Zimdahl, D. Pavon, 
Phys. Lett. B {\bf 521} (2001) 133-138 [e-Print:astro-ph/0105479].

\bibitem{quint-7} L. P. Chimento, A. S. Jakubi, D. Pavon, W. Zimdahl, 
Phys. Rev. D {\bf 67} (2003) 083513 [e-Print:astro-ph/0303145].


\bibitem{Caldwell:2005tm}
R.~Caldwell and E.~V.~Linder,
Phys. Rev. Lett. \textbf{95} (2005), 141301
doi:10.1103/PhysRevLett.95.141301
[arXiv:astro-ph/0505494].

\bibitem{phantom-1} R. R. Caldwell, 
Phys. Lett. B {\bf 545} (2002) 23-29 [e-Print:astro-ph/9908168].

\bibitem{phantom-2} R. R. Caldwell, M. Kamionkowski, N. N. Weinberg, 
Phys. Rev. Lett. {\bf 91} (2003) 071301 [e-Print:astro-ph/0302506].

\bibitem{phantom-3} E. Elizalde, S. Nojiri, S. D. Odintsov, 
Phys. Rev. D {\bf 70} (2004) 043539 [e-Print:hep-th/0405034].

\bibitem{phantom-4} S. Nojiri, S. D. Odintsov, S. Tsujikawa, 
Phys. Rev. D {\bf 71} (2005) 063004 [e-Print:hep-th/0501025].

\bibitem{Kujat:2006vj}
Kujat, Jens and Scherrer, Robert J. and Sen, A.A.
Phys. Rev. D, 74, 083501, 2006
[astro-ph/0606735].



\bibitem{Ludwick:2017tox}
K.~J.~Ludwick,
Mod. Phys. Lett. A \textbf{32} (2017) no.28, 1730025
doi:10.1142/S0217732317300257
[arXiv:1708.06981].

\bibitem{k-1} T. Chiba, 
Phys. Rev. D {\bf 66} (2002) 063514 [e-Print:astro-ph/0206298].

\bibitem{k-2} M. Malquarti, E. J. Copeland, A. R. Liddle, M. Trodden, 
Phys. Rev. D {\bf 67} (2003) 123503 [e-Print:astro-ph/0302279].

\bibitem{k-3} R. J. Scherrer, 
Phys. Rev. Lett. {\bf 93} (2004) 011301 [e-Print:astro-ph/0402316].

\bibitem{k-4} L. P. Chimento, A. Feinstein, 
Mod. Phys. Lett. A {\bf 19} (2004) 761-768 [e-Print:astro-ph/0305007].

\bibitem{qton-1} Z. K. Guo, Y. S. Piao, X. M. Zhang, Y. Z. Zhang, 
Phys. Lett. B {\bf 608} (2005) 177-182 [e-Print:astro-ph/0410654].

\bibitem{qton-2} H. Wei, R. G. Cai, D. F. Zeng, 
Class. Quant. Grav. {\bf 22} (2005) 3189-3202 [e-Print:hep-th/0501160].

\bibitem{qton-3} B. Feng, M. Li, Y. S. Piao, X. M. Zhang, 
Phys. Lett. B {\bf 634} (2006) 101-105 [e-Print:astro-ph/0407432].

\bibitem{qton-4} W. Zhao, 
Phys. Rev. D {\bf 73} (2006) 123509 [e-Print:astro-ph/0604460].

\bibitem{Cai:2009zp}
  Y.~F.~Cai, E.~N.~Saridakis, M.~R.~Setare and J.~Q.~Xia,
  Phys.\ Rept.\  {\bf 493} (2010) 1
  doi:10.1016/j.physrep.2010.04.001
  [arXiv:0909.2776].

\bibitem{coinc-p} L.P. Chimento, A.S. Jakubi, D. Pavon, W. Zimdahl, Phys. Rev. D {\bf 67} (2003) 083513 [arxiv: astro-ph/0303145]; R.G. Cai, A. Wang, JCAP {\bf 03} (2005) 002 [arxiv: hep-th/0411025]; B. Hu, Y. Ling, Phys. Rev. D {\bf 73} (2006) 123510 [arxiv: hep-th/0601093]; S. Dodelson, M. Kaplinghat, E. Stewart, Phys. Rev. Lett. {\bf 85} (2000) 5276-5279 [arxiv:astro-ph/0002360].

\bibitem{1} G. Alestas, L. Kazantzidis, L. Perivolaropoulos, ``$H_0$ Tension, Phantom Dark Energy and Cosmological Parameter Degeneracies'', 
[e-Print:2004.08363].

\bibitem{2} A. Bouali, I. Albarran, M. Bouhmadi-L\'opez, T. Ouali, ``Cosmological constraints of phantom dark energy models'', Phys. Dark Univ. {\bf 26} (2019) 100391 [e-Print:1905.07304].

\bibitem{3} D. Perkovic, H. Stefancic, ``Dark sector unifications: dark matter-phantom energy, dark matter - constant $w$ dark energy, dark matter-dark energy-dark matter'', Phys. Lett. B {\bf 797} (2019) 134806 [e-Print:1902.05365].

\bibitem{4} K.J. Ludwick, ``Viability of phantom dark energy as a quantum field in first-order perturbation theory of FLRW spacetime'', Phys. Rev. D {\bf 98} (2018) 043519 [e-Print:1804.02987].

\bibitem{5} N. Roy, N. Bhadra, ``Dynamical systems analysis of phantom dark energy models'', JCAP {\bf 06} (2018) 002 [e-Print:1710.05968].

\bibitem{Ratra:1987rm}
B.~Ratra and P.~Peebles,
Phys. Rev. D \textbf{37} (1988), 3406
doi:10.1103/PhysRevD.37.3406

\bibitem{Ferreira:1997hj}
P.~G.~Ferreira and M.~Joyce,
Phys. Rev. D \textbf{58} (1998), 023503
doi:10.1103/PhysRevD.58.023503
[arXiv:astro-ph/9711102].

\bibitem{Brax:1999gp}
P.~Brax and J.~Martin,
Phys. Lett. B \textbf{468} (1999), 40-45
doi:10.1016/S0370-2693(99)01209-5
[arXiv:astro-ph/9905040].

\bibitem{Chang:2016aex}
H.~Y.~Chang and R.~J.~Scherrer,
[arXiv:1608.03291].

\bibitem{Scherrer:2008be}
R.~J.~Scherrer and A.~Sen,
Phys. Rev. D \textbf{78} (2008), 067303
doi:10.1103/PhysRevD.78.067303
[arXiv:0808.1880].

\bibitem{Dutta:2009dr}
S.~Dutta and R.~J.~Scherrer,
Phys. Lett. B \textbf{676} (2009), 12-15
doi:10.1016/j.physletb.2009.04.072
[arXiv:0902.1004].

\bibitem{Dutta:2008qn}
S.~Dutta and R.~J.~Scherrer,
Phys. Rev. D \textbf{78} (2008), 123525
doi:10.1103/PhysRevD.78.123525
[arXiv:0809.4441].

\bibitem{Avsajanishvili:2017zoj}
O.~Avsajanishvili, Y.~Huang, L.~Samushia and T.~Kahniashvili,
Eur. Phys. J. C \textbf{78} (2018) no.9, 773
doi:10.1140/epjc/s10052-018-6233-y
[arXiv:1711.11465].

\bibitem{Durrive:2018quo}
J.~B.~Durrive, J.~Ooba, K.~Ichiki and N.~Sugiyama,
Phys. Rev. D \textbf{97} (2018) no.4, 043503
doi:10.1103/PhysRevD.97.043503
[arXiv:1801.09446].

\bibitem{Lonappan:2017lzt}
A.~I.~Lonappan, S.~Kumar, Ruchika, B.~R.~Dinda and A.~A.~Sen,
Phys. Rev. D \textbf{97} (2018) no.4, 043524
doi:10.1103/PhysRevD.97.043524
[arXiv:1707.00603].

\bibitem{Roy:2018nce}
N.~Roy, A.~X.~Gonzalez-Morales and L.~A.~Urena-Lopez,
Phys. Rev. D \textbf{98} (2018) no.6, 063530
doi:10.1103/PhysRevD.98.063530
[arXiv:1803.09204].

\bibitem{Tosone:2018qei}
F.~Tosone, B.~S.~Haridasu, V.~V.~Luković and N.~Vittorio,
Phys. Rev. D \textbf{99} (2019) no.4, 043503
doi:10.1103/PhysRevD.99.043503
[arXiv:1811.05434].

\bibitem{Yang:2018xah}
W.~Yang, M.~Shahalam, B.~Pal, S.~Pan and A.~Wang,
Phys. Rev. D \textbf{100} (2019) no.2, 023522
doi:10.1103/PhysRevD.100.023522
[arXiv:1810.08586].

\bibitem{Lazkoz:2007mx}
Lazkoz, Ruth and Leon, Genly and Quiros, Israel,
Phys. Lett. B", 649, 103--110, 2007.
doi:10.1016/j.physletb.2007.03.060
[arXiv:0701353].

\bibitem{Scherrer:2007pu}
Scherrer, Robert J. and Sen, A.A.,
Phys. Rev. D, 77, 083515, 2008
doi:10.1103/PhysRevD.77.083515
[arXiv:0712.3450].

\bibitem{Matos:2008ag}
Matos, Tonatiuh and Vazquez-Gonzalez, Alberto and Magana, Juan.
Mon. Not. Roy. Astron. Soc. 393, 1359--1369, 2009.
doi:10.1111/j.1365-2966.2008.13957.x,
[arXiv:0806.0683].

\bibitem{Gonzalez:2008wa}
Gonzalez, Tame and Matos, Tonatiuh and Quiros, Israel and Vazquez-Gonzalez, Alberto,
Phys. Lett. B, 676, 161--167, 2009
doi: "10.1016/j.physletb.2009.04.080"
[arXiv:0812.1734].

\bibitem{Chiba:2005tj}
Chiba, Takeshi
Phys. Rev. D, 73, 063501, 2006
doi: 10.1103/PhysRevD.80.129901
[arXiv:astro-ph/0510598].

\bibitem{SimpleMC}
Temporarily stored at: {\rm https://github.com/ja-vazquez/SimpleMC}

\bibitem{Aubourg:2014yra}
Aubourg, Eric and {\it et.al}
Phys. Rev. D, 92, 12, 123516, 2015.
doi: 10.1103/PhysRevD.92.123516
[arXiv:1411.1074].  

\bibitem{Sahni:1999qe}
Sahni, Varun and Wang, Li-Min,
Phys. Rev. D, 62, 103517, 2000
doi: 10.1103/PhysRevD.62.103517
[arXiv:astro-ph/9910097].

\bibitem{UrenaLopez:2000aj}
Urena-Lopez, L.Arturo and Matos, Tonatiuh
Phys. Rev. D, 62, 081302, 2000
doi: 10.1103/PhysRevD.62.081302
[arXiv:astro-ph/0003364].




\bibitem{Padilla:2019fju}
Padilla, Luis E. and V\'azquez, J. Alberto and Matos, Tonatiuh and Germ\'an, Gabriel,
JCAP, 05, 056, 2019
doi: 10.1088/1475-7516/2019/05/056
[arXiv:1901.00947].

\bibitem{Matos:2009hf}
Matos, Tonatiuh and Luevano, Jose-Ruben and Quiros, Israel and Urena-Lopez, L.Arturo and Vazquez, Jose Alberto
Phys. Rev. D, 80, 123521, 2009
doi: 10.1103/PhysRevD.80.123521
[arXiv:0906.0396].



\bibitem{Ross:2016hyb}
Ross, Graham G. and German, Gabriel and Vazquez, J. Alberto
JHEP, 05, 010, 2016
doi: 10.1007/JHEP05(2016)010
[arXiv:1601.03221].

\bibitem{German:2019aoj}
Germ\'an, Gabriel and Hidalgo, Juan Carlos and Linares Cedeño, Francisco X. and Montiel, Ariadna and V\'azquez, J. Alberto",
Phys. Rev. D, 101, 2, 023507, 2020
doi: 10.1103/PhysRevD.101.023507
[arXiv:1909.02019].


\bibitem{Speagle:2020}
Speagle, Joshua S.,
Mon. Not. Roy. Astron. Soc. \textbf{493} (2020) no.3, 3132
doi: 10.1093/mnras/staa278
[arXiv:1904.02180].

\bibitem{Skilling:2004}
Skilling, John,
AIP Conference Proceedings \textbf{735} (2004) no.1, 395
doi: 10.1063/1.1835238

\bibitem{Skilling:2006}
Skilling, John,
Bayesian Anal. \textbf{12} (2006) no.4, 855
doi: 10.1214/06-BA127



\bibitem{Vazquez:2011xa}
Vazquez, J.Alberto and Lasenby, A.N. and Bridges, M. and Hobson, M.P.,
Mon. Not. Roy. Astron. Soc., 422, 1948--1956, 2012    
doi: 10.1111/j.1365-2966.2012.20606.x
[arXiv:1103.4619].


\bibitem{Vazquez:2012ux}
Vazquez, J.Alberto and Bridges, M. and Hobson, M.P. and Lasenby, A.N.",
JCAP, 06, 006, 2012
doi: 10.1088/1475-7516/2012/06/006
[arXiv:1203.1252].

\bibitem{Padilla:2019mgi}
Padilla, Luis E. and Tellez, Luis O. and Escamilla, Luis A. and Vazquez, J. Alberto,
[arXiv:1903.11127].




\bibitem{Alam:2016hwk}
Alam, Shadab and others (BOSS),
Mon. Not. Roy. Astron. Soc., 470, 3, 2017
doi: 10.1093/mnras/stx721
[arXiv:1607.03155].



\bibitem{Blomqvist:2019rah}
Blomqvist, Michael and others
Astron. Astrophys., 629, A86, 2019
doi: 10.1051/0004-6361/201935641
[arXiv:1904.03430].


\bibitem{Ata:2017dya}
Ata, Metin and others
Mon. Not. Roy. Astron. Soc., 473, 4, 2018
doi: 10.1093/mnras/stx2630
[arXiv:1705.06373].    



\bibitem{Agathe:2019vsu}
de Sainte Agathe, Victoria and others
Astron. Astrophys., 629, A85, 2019
doi: 10.1051/0004-6361/201935638
[arXiv:1904.03400].    



\bibitem{Beutler2011}
Beutler, Florian and Blake, Chris and others.
Mon. Not. Roy. Astron. Soc., 416, 4, 2011
doi: 10.1111/j.1365-2966.2011.19250.x
[arXiv:1106.3366].


\bibitem{Anderson:2013zyy}
Anderson, Lauren and others
Mon. Not. Roy. Astron. Soc., 441, 1, 2014
doi: 10.1093/mnras/stu523
[arXiv:1312.4877].    
    


\bibitem{Gomez-Valent:2018hwc}
G\'omez-Valent, Adri\`{a} and Amendola, Luca,
JCAP, 04, 051, 2018
doi: 10.1088/1475-7516/2018/04/051
[arXiv:1802.01505].


\bibitem{Scolnic:2017caz}
Scolnic, D.M. and others
Astrophys. J., 859, 2, 101, 2018
doi: 10.3847/1538-4357/aab9bb
[arXiv:1710.00845].    

\bibitem{Vagnozzi:2019ezj}
S.~Vagnozzi,
Phys. Rev. D \textbf{102} (2020) no.2, 023518
doi:10.1103/PhysRevD.102.023518
[arXiv:1907.07569 [astro-ph.CO]].

\bibitem{Handley}
Will Handley, fgivenx:  
A Python package for functional posterior plotting, 
[arXiv:1908.01711].

\end{thebibliography}
\end{document}